\documentclass[%
 reprint,
nofootinbib,
 amsmath,amssymb,
 aps,
]{revtex4-2}

\usepackage{graphicx}
\usepackage{dcolumn}
\usepackage{bm}

\usepackage{feynmp}
\usepackage{tikz-feynman}
\tikzfeynmanset{compat=1.1.0}
\usetikzlibrary{shapes,arrows,positioning,automata,backgrounds,calc,er,patterns}

\usepackage{amsmath}
\usepackage{amssymb}
\usepackage[colorlinks=true,citecolor=magenta,urlcolor=cyan]{hyperref}
\usepackage{multirow}

\usepackage{lineno}

\def\antiq{\bar{q}}
\def\q{q}
\def\qp{q'}
\def\h{h}
\def\sh{s_{h}}
\def\qbar{\bar{q}}
\def\qpbar{\bar{q}'}
\def\hbar{H}

\def\eh{\varepsilon_{\h}}

\def\sigmaZq{\sigma_z}

\def\Iden{1}

\def\GeV{\rm{GeV}}

\def\xu{\hat{\textbf{x}}}
\def\yu{\hat{\textbf{y}}}
\def\zu{\hat{\textbf{z}}}

\def\p{p}
\def\k{k}
\def\kp{k'}
\def\pt{\textbf{p}_{\rm T}}

\def\kT{\textbf{k}_{\rm T}}
\def\kTkT{\rm{k}^2_{\rm T}}

\def\kt{\textbf{k}_{\rm T}}
\def\kpt{\textbf{k}_{\rm T}}
\def\ktkt{\textbf{k}^2_{\rm T}}

\def\kptkpt{\textbf{k}^2_{2\rm T}}
\def\ptpt{\textbf{p}^2_{\rm T}}

\def\kptabs{\rm{k}_{2\rm T}}

\def\mh{M_h}
\def\bL{b_{\rm L}}
\def\bT{b_{\rm T}}
\def\fT{f_{\rm T}}
\def\sigmaT{\boldsymbol{\sigma}_{\rm T}}

\def\i{\rm i}
\def\sigmaqT{\boldsymbol{\sigma}_{\rm T}}

\def\GL{G_{\rm L}}
\def\GT{G_{\rm T}}
\def\V{\textbf{V}}
\def\VL{V_{\rm L}}
\def\VT{\textbf{V}_{\rm T}}
\def\fVM{f_{\rm VM}}
\def\fL{f_{\rm L}}

\def\Gammaq{\Gamma}

\def\PS{\rm{PS}}
\def\VM{\rm{VM}}

\def\Tr{\rm{Tr}}

\def\Sq{\textbf{S}_{\q}}
\def\SqT{\textbf{S}_{q,\T}}
\def\SqL{S_{q,\L}}

\def\T{\textbf{T}}

\def\be{\begin{equation}}
\def\ee{\end{equation}}

\def\eg{{\sl e.g.}}

\def\3p0{string+${}^3P_0$}
\def\Sq{\textbf{S}_q}
\def\SB{\textbf{S}_B}
\def\SqT{\textbf{S}_{q,\rm T}}
\def\SqL{S_{q,\rm z}}

\def\Sqp{\textbf{S}_{q'}}

\def\SbTone{\textbf{S}^{(1)}_{B,\rm T}}
\def\SbLone{S^{(1)}_{B,\rm L}}
\def\SbTzero{\textbf{S}^{(0)}_{B,\rm T}}
\def\SbLzero{S^{(0)}_{B,\rm L}}

\def\kt{\textbf{k}_{\T}}

\def\pt{\textbf{p}_{\T}}
\def\ptpt{\textbf{p}^2_{\T}}
\def\kpt{\textbf{k}'_{\T}}

\def\kptabs{\rm{k}'_{\T}}
\def\kptkpt{\textbf{k}^{'2}_{\T}}
\def\kpx{k'_x}
\def\kpxkpx{k^{'2}_x}
\def\kpy{k'_y}
\def\kpykpy{k^{'2}_y}
\def\kx{k_x}

\def\ky{k_y}

\def\T{\rm{T}}
\def\fT{f_{\T}}
\def\Tr{\rm{Tr}}
\def\qA{q_{\rm{A}}}

\def\qBb{\bar{q}_{\rm{B}}}

\def\epsB{\varepsilon_B}
\def\epsBbar{\varepsilon_{\Bbar}}
\def\aq{a_q}
\def\aDbar{a_{D}}
\def\aD{a_D}
\def\bL{b_{\rm L}}
\def\MB{M_B}
\def\MBbar{M_{\Bbar}}
\def\sigmaT{\boldsymbol{\sigma}_{\rm T}}

\def\uhat{\hat{u}}

\def\xu{\hat{\bf{x}}}
\def\yu{\hat{\bf{y}}}
\def\zu{\hat{\bf{z}}}

\def\Sqz{S_{q,z}}

\def\i{\rm i}
\def\Im{\rm Im}
\def\Re{\rm Re}
\def\Sqq{\textbf{S}_{\rm qq}}

\def\phat{\hat{\textbf{p}}}

\def\khat{\boldsymbol{\hat{k}}}
\def\k{k}

\def\RT{\textbf{k}_{\rm T}}

\def\qbar{\bar{q}}
\def\R{\textbf{k}}
\def\Rx{k_{x}}
\def\RxRx{k_{x}^2}
\def\Ry{k_{y}}
\def\RyRy{k^2_{y}}
\def\Rz{k_{z}}
\def\RzRz{k_{z}^2}

\def\i{\rm i}

\def\kv{{\bf{k}}}
\def\T{_{\rm T}}

\def\be{\begin{equation}}
\def\ee{\end{equation}}

\def\qq{qq}

\def\qqbar{\overline{qq}}

\def\DeltaD{\Delta_{qq}}

\def\Q{Q}
\def\Qp{Q'}
\def\qbarp{\bar{q}'}
\def\aq{a_q}
\def\aQ{a_{\Q}}
\def\aQp{a_{\Qp}}
\def\aqp{a_{q'}}
\def\aqq{a_{\qq}}
\def\aqqbar{a_{\qqbar}}
\def\Sh{S_h}
\def\t{t}
\def\eps{\varepsilon}
\def\B{\rm B}
\def\kbf{\textbf{k}}
\def\muqq{\mu_{\qq}}
\def\mqq{m_{\qq}}
\def\SQ{S_{\Q}}

\def\Mcov{\hat{\mathcal{M}}_{cov.}}
\def\SBp{\textbf{S}_{B'}}
\def\Mhat{\hat{\mathcal{M}}}

\def\epsbf{\boldsymbol{\phi}}

\def\sigmabf{\boldsymbol{\sigma}}
\def\EM{\rm{EM}}
\def\NL{\rm{NL}}
\def\RD{\rm{WR}}

\def\rhocheck{\check{\rho}}
\def\Pp{\textbf{P}'}
\def\Pphat{\boldsymbol{\hat{P}}'}

\def\aC{a_{C,h}}
\def\ptabs{p_{\rm T}}
\def\ptabsptabs{p_{\rm T}^2}
\def\pthat{\hat{\textbf{p}}_{\rm T}}
\def\Lambdabar{\bar{\Lambda}}

\def\Shcheck{\check{\textbf{S}}_B}
\def\phipp{\phi_{\Pphat}}
\def\thetapp{\theta_{\Pphat}}
\def\SY{\textbf{S}_Y}
\def\SBpcheck{\check{\textbf{S}}_{B'}}
\def\SYcheck{\check{\textbf{S}}_{Y}}
\def\Bbar{\bar{B}}
\def\SBbar{\textbf{S}_{\Bbar}}
\def\SBbarT{\textbf{S}_{\Bbar,\rm T}}
\def\SBbarL{S_{\Bbar,\rm L}}
\def\SBbarcheck{\check{\textbf{S}}_{\Bbar}}
\def\SqqbarT{\textbf{S}_{\qqbar,\rm T}}
\def\SqqbarL{S_{\qqbar,\rm L}}
\def\Sqqbar{\textbf{S}_{\qqbar}}
\def\Hone{\textrm{H}_{\rm 1}}
\def\Htwo{\textrm{H}_{\rm 2}}

\begin{document}

\preprint{APS/123-QED}
\title{A model for baryon production in spin-dependent string fragmentation}
\author{A. Kerbizi$^{a}$}\email{albi.kerbizi@ts.infn.it}
\author{X. Artru$^{b}$} \email{xavier.artru@orange.fr}
\affiliation{
$^{a}$Department of physics, Lund University, Box 118, 221 00 Lund, Sweden, \\
INFN Sezione di Trieste,Via Valerio 2, 34127 Trieste, Italy\\ 
$^{b}$Universit\'e de Lyon, Institut de Physique des deux Infinis (IP2I Lyon), Universit\'e Lyon 1 and CNRS,\\
 4 rue Enrico Fermi, F-69622 Villeurbanne, France
}%

\date{\today}

\begin{abstract}
We introduce spin-1/2 baryons in the string+${}^3P_0$ model of hadronization, previously restricted to the production of pseudoscalar and vector mesons. Baryons are modeled as quark-diquark bound states and baryon production is described by the tunneling of diquark-antidiquark pairs at string breaking points. Diquarks can be scalar or pseudovector, the latter being produced in the relative ${}^5D_0$ state. Introducing the quark-baryon-diquark coupling, the relevant splitting amplitudes for the emission of baryons are constructed and used to explore analytically the model predictions. We find a Collins effect for baryon production in the fragmentations of transversely polarized quarks or diquarks as well as a baryon spontaneous transverse polarization. The hadronic decays of polarized hyperons are also included in the model, which is presented in a form suitable for the implementation in a Monte Carlo event generator.
 \end{abstract}

\maketitle


\section{Introduction}
Quark spin effects in hadronization, the process of conversion of quarks and gluons in hadrons, have been shown to be non negligible in different processes such as proton-proton scattering, semi-inclusive deep inelastic scattering (SIDIS) and $e^+e^-$ annihilation to hadrons. They are commonly described in quantum chromodynamics (QCD) by the fragmentation functions (FFs), which encode the underlying non-perturbative and poorly known physics mechanisms.

An interesting phenomenon is fragmentation $q^{\uparrow}\rightarrow h+X$ of a transversely polarized quark $q^{\uparrow}$ in the hadron $h$, known as the Collins effect \cite{Collins:1992kk}. It is described by the FF
\begin{eqnarray}\label{eq:Collins FF}
\nonumber D_{q^{\uparrow}\rightarrow h +X}(z,\pt)=D_{1q}^{h}(z,\ptpt) \,\bigg[1+\aC\,\SqT\cdot(\khat\times \pthat)\bigg],\\
\end{eqnarray}
where $D_{1q}^h$ is the spin averaged FF depending on the fraction $z$ of the forward lightcone momentum of $q$ taken by $h$ and on the transverse momentum $\pt$ of $h$ with respect to the quark momentum $\kv=|\kv|\khat$. If the quark transverse polarization $\SqT$ is different from zero, the mixed product in the second term in Eq. (\ref{eq:Collins FF}) produces a modulation in the distribution of the azimuthal angle of the hadron. The amplitude of the modulation is given by the Collins analysing power $\aC(z,\ptabs)$, which depends on the \textit{Collins FF} \cite{Collins:1992kk}. The Collins effect has been measured to be non vanishing in semi-inclusive deep inelastic scattering (SIDIS) with a transversely polarized proton target \cite{HERMES:2020ifk,COMPASS:2014bze} and in $e^+e^-$ annihilation to hadrons \cite{Belle:2008fdv,Belle:2019nve, BaBar:2013jdt,BaBar:2015mcn,BESIII:2015fyw}.

If pairs of hadrons $h_1h_2$ are measured among the products of the fragmentation of the same quark in the process $q^{\uparrow}\rightarrow h_1h_2+X$, a modulation in the relative momentum of the pair, known as the dihadron production asymmetry, shows up \cite{Collins:1993kq,Bianconi:1999cd}. The effect is described by the \textit{interference FF} and has been measured to be non-vanishing in SIDIS with a transversely polarized proton target \cite{HERMES:2010mmo,COMPASS:2014ysd} and in $e^+e^-$ annihilation to hadrons \cite{Belle:2011cur}.

The Collins FF and the IFF are particularly important as they provide access to the partonic transverse spin structure of the nucleons and have been extracted by different groups using SIDIS and $e^+e^-$ data (for a review see, \textit{e.g.}, \cite{Anselmino:2020vlp}).

Spin effects show up also in the fragmentation of unpolarized quarks. An example is the \textit{spontaneous} transverse polarization of $\Lambda$ and $\bar{\Lambda}$ hyperons, \textit{i.e.} the polarization along the vector perpendicular to the hyperon production plane. The effect was measured to be non-vanishing for the first time in the 70's in high energy hadronic collisions \cite{Panagiotou:1989sv}. More recently, it was measured by the HERMES experiment in deep inelastic scattering by the reaction $lN\rightarrow YX$, with $Y=\Lambda,\Lambdabar$ \cite{HERMES:2014fmx}, and in $e^+e^-$ annihilation by the BELLE experiment via the reactions $e^+e^-\rightarrow Y + X$ and $e^+e^-\rightarrow Y+\pi(K)+X$, where the hyperon and the meson $\pi$ or $K$ are produced in opposite hemispheres in the center of mass system of the event \cite{Belle:2018ttu}.

After the first measurements of the spontaneous polarization of hyperons, different models were put forward to explain the observed effects (for a review see , \textit{e.g.}, Ref. \cite{Felix:1999tf}). A description of the $\Lambda$ polarization was given by the Lund group \cite{Andersson:1979wj} in the frame of the Lund Model (LM) of string fragmentation \cite{Andersson:1983ia}. The polarization was explained as a consequence of the correlation between the spins and the transverse momenta of $q\qbar$ pairs produced at string breakings in the ${}^3P_0$ state, \textit{i.e.} with orbital angular momentum $L=1$ and total spin $S=1$ such that $\textbf{J}=\textbf{L}+\textbf{S}$ vanishes.

The current QCD description of the spontaneous polarization of a spin 1/2-baryon $B^{\uparrow}$ produced in the fragmentation $q\rightarrow \B^{\uparrow} + X$ is given in terms of the \textit{polarizing FF} \cite{Mulders:1995dh,Anselmino:2000vs}. The complete FF takes the form
\begin{eqnarray}\label{eq:polarizing FF}
\nonumber    D_{q\rightarrow B^{\uparrow}+X}(z,\pt)=\frac{D_{1q}^B(z,\ptpt)}{2}\,\bigg[1+a_B\,\Shcheck\cdot(\khat\times \pthat)\bigg],\\
\end{eqnarray}
where $\Shcheck$ is a chosen direction for the polarization vector of $B$. The quantity $a_B$ is related to the polarizing FF and it gives the magnitude of the polarization of $B$. The predicted polarization of $B$ is directed along $\khat\times\phat$, \textit{i.e.} along the vector perpendicular to the production plane of $B$. An extraction of the polarizing FFs for $\Lambda$ and $\Lambdabar$ has been performed for proton-proton data in Ref. \cite{Anselmino:2000vs} and more recently for $e^+e^-$ data from BELLE in Refs. \cite{DAlesio:2020wjq,Callos:2020qtu,DAlesio:2022brl}.

A different approach to tackle the quark-spin dependence of hadronization is by modeling the underlying physics and by implementing the model in a Monte Carlo event generator (MCEG).
This is the approach that we have followed by developing the string+${}^3P_0$ model of polarized hadronization \cite{Kerbizi:2018qpp,Kerbizi:2019ubp,Kerbizi:2021M20}. The string+${}^3P_0$ model is a recursive quantum mechanical model that accounts for the systematic propagation of the spin information along the fragmentation chain. It is inspired by the multiperipheral model \cite{Amati:1962nv}, and implements the string fragmentation dynamics of the LM and the assumption that $q\qbar$ pairs at string breakings are produced in the ${}^3P_0$ relative state. This is achieved by introducing a spin-dependent $2\times 2$ quark propagator proportional to the matrix $\Delta_q=\mu_q+\sigma_z\sigmabf\cdot\kt$, where $\mu_q$ is a free complex parameter referred to as the complex mass and $\kt$ is the transverse momentum of the fragmenting quark with respect to the string axis, and $2\times 2$ vertex matrices $\Gamma_{h}$. The model was implemented in the Pythia 8 MCEG \cite{Bierlich:2022pfr} via the StringSpinner package \cite{Kerbizi:2021StringSpinner} and was shown to reproduce the Collins asymmetries and the dihadron production asymmetries in SIDIS \cite{Kerbizi:2023cde} and the Collins asymmetries in $e^+e^-$ \cite{Kerbizi:2023luv,Kerbizi:2024vpd}.

The string+${}^3P_0$ model is however restricted to the production of pseudoscalar (PS) mesons and vector mesons (VMs) in the final states but lacks the production of baryons. The introduction of the production and decay of polarized spin-1/2 baryons in the string+${}^3P_0$ model of Ref. \cite{Kerbizi:2021M20} is the goal of this work\footnote{The heavier spin-3/2 baryon states are less frequently produced in hadronization and their inclusion in the model is not compelling. It is left for future work.}.
We describe baryon production by allowing diquark-antidiquark pairs $(\qq)-(\qqbar)$ to be produced at the string breakings, as suggested in the spinless LM in Ref. \cite{Andersson:1981ce} and implemented in the Pythia MCEG. We distinguish between scalar $(\qq)_0$ and pseudovector (PV) $(\qq)_1$ diquarks. In the latter case the diquark-antidiquark pair should be created in the ${}^1S_0$ or in the ${}^5D_0$ state, \textit{i.e.}, with the vacuum quantum numbers, by analogy with the ${}^3P_0$ state of the $q\qbar$ pair creation. In this paper we introduce a \textit{diquark propagator} $\DeltaD$ in spin space based on the dominance of the ${}^5D_0$ state. We also introduce the spin matrices $\Gamma_B$ for the quark-baryon-diquark vertices. The $\DeltaD$ and $\Gamma_B$ matrices are the basic objects of our model. From them we build \textit{splitting matrix-functions} $T=F\times\DeltaD\times \Gamma_B$, where $F$ is a function of relevant momenta, baryon species and diquark species. The $T$ matrices describe the elementary splittings $q\rightarrow B+(\qqbar)$ and $(\qqbar)\rightarrow \Bar{B}+q'$ in spin and momentum space. They allow to study qualitatively the model predictions as well as to implement the model in a MCEG. The MCEG implementation of the model will be presented in a separate work.

The paper is organized as follows. The splitting matrices of the extended string+${}^3P_0$ model with meson and baryon production are constructed in Sec. \ref{sec:the new model}. In Sec. \ref{sec:probability distribution and collins effect} we study the probability distribution of the produced baryons, in particular the Collins effect for baryon production. The predicted spin states of the produced baryons, among which the spontaneous polarization, are described in Sec. \ref{sec:polarization}. The subsequent decays of the polarized unstable hyperons are described in Sec. \ref{sec:hyperon decays}. Section \ref{sec:propagation of spin information} describes the rules for the propagation of the spin information along the fragmentation chain. We give the conclusions in Sec. \ref{sec:conclusions}.

\section{The extended string+${}^3P_0$ model with meson and baryon production}\label{sec:the new model}
We start by considering the hadronization
\begin{equation}\label{eq:qAqBbar->h1h2..}
\q_{A}\antiq_{B}\rightarrow h_1\,h_2\,\dots,
\end{equation}
of the quark-antiquark pair $\q_{A}\qbar_{B}$ to the final state hadrons $h_1,h_2,\dots$. The hadronization is viewed as the breaking of the string stretched between $\q_A$ and $\qbar_B$, in the center of mass system of the pair, which occurs by the iteration of the elementary splittings $\q\rightarrow \h + \qp$, where $\q$ is the fragmenting quark, $\h$ is the emitted hadron and $\qp$ is the leftover quark. In the string+${}^3P_0$ model of Ref. \cite{Kerbizi:2021M20}, the hadron $h=q\qbarp$ was restricted to be either a PS meson or a VM.

To generalize the string+${}^3P_0$ model to include the production of polarized baryons, which are modeled as quark-diquark bound states, it is necessary to describe the elementary splittings $q\rightarrow B+(\qqbar)$ and $(\qqbar)\rightarrow \Bbar+\qp$, where $B$ ($\Bbar$) refers to the emitted baryon (antibaryon) and $(\qq)$ [$(\qqbar)$] refer to the fragmenting diquark (leftover antidiquark). We will thus consider the more general splitting
\begin{eqnarray}\label{eq:splittings}
 \Q\rightarrow h +\Qp =
 \begin{cases}
    \q\rightarrow M +\qp \\
    \q\rightarrow B + (\qqbar)\\
    (\qqbar)\rightarrow \Bbar + \qp
    \end{cases},
\end{eqnarray}
where the fragmenting particle $Q$ can be a quark ($Q=q$) or an antidiquark $Q=(\qqbar)$, the hadron $h$ can be a PS meson or a VM ($h=M$), a baryon ($h=B$) or an antibaryon ($h=\Bbar$), and the leftover particle $\Qp$ can be a quark $(\Qp=\qp)$ or an antidiquark $[\Qp=(\qqbar)]$.

We will describe the hadronization of the $\qA\qBb$ pair the center of mass system of the pair, which is referred to as the string frame. In this frame we introduce the axes $\zu=\kv_{\rm A}/|\kv_{\rm A}|$, $\yu=\hat{\textbf{u}}\times \zu$ and $\xu=\yu\times \zu$, where $\kv_{\rm A}$ is the momentum of $\qA$ and $\hat{\textbf{u}}$ is a reference vector, \eg{}, the electron beam direction in $e^+e^-$ annihilation. The string axis is along the $\zu$ axis and it defines the longitudinal direction.

For each splitting we indicate by $\k$, $\p$ and $\kp=k-p$ the four momenta of $\Q$, $h$ and $\Qp$, respectively. The four-momentum of $\h$ is parametrized in terms of the longitudinal splitting variable $Z=\p^+/\k^+$, the transverse momentum (with respect to the string axis) $\pt=(p_x,p_y)$ and the transverse energy $\eh=\sqrt{\mh^2+\ptpt}$, $\mh$ being the mass of $h$. The lightcone components for a generic four-vector $v$ are defined as $v^{\pm} = v^0\pm v^{z}$. We have $\pt=\kt-\kpt$, $\kt$ and $\kpt$ being the transverse momenta of $\Q$ and $\Qp$, respectively.

The description of the elementary splitting $\Q\rightarrow h+\Qp$ in momentum and spin space is achieved by the means of a splitting matrix $T_{\Qp,h,\Q}$. 
We build such matrix using the LM of string fragmentation with the addition of spin matrices $\Delta$ (quark or diquark propagators) and $\Gamma$ (coupling matrices). The general expression for $T_{\Qp,h,\Q}$ is introduced in Sec. \ref{sec:new T for quarks}, while the expressions for $\Gamma$ and $\Delta$ are given in Sec. \ref{sec:Gamma} and Sec. \ref{sec:propagator}, respectively.

\begin{figure*}[tb]
\centering
\includegraphics[width=1.0\textwidth]{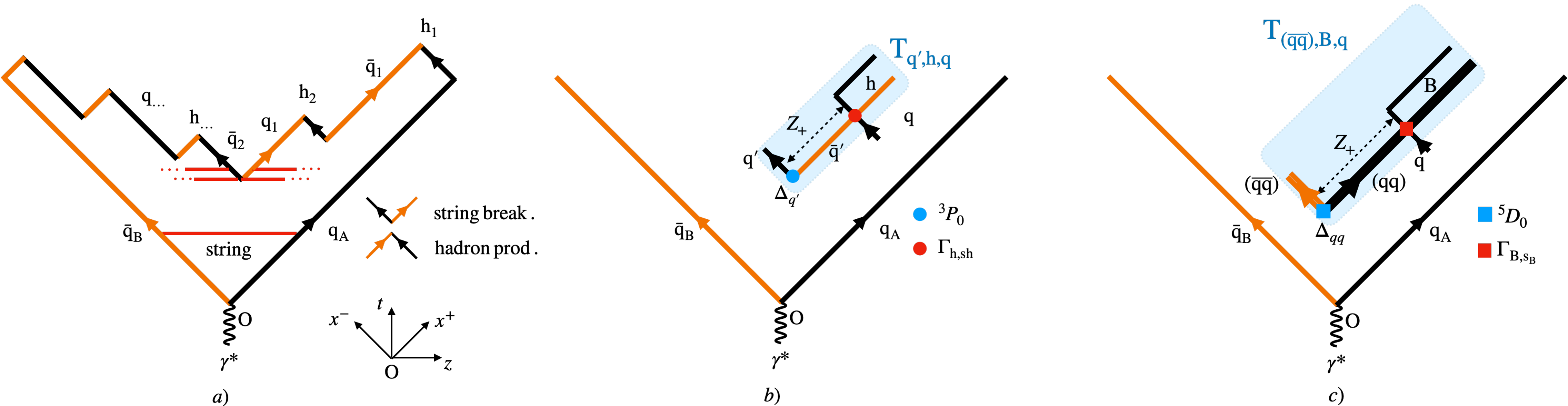}
\caption{Space-time picture of spin-less string fragmentation (a), and of string breaking via the tunneling of a quark-antiquark pair (b) or a diquark-antidiquark pair (c). The shaded areas enclose the elements that define the quark splitting matrices for the emission of a meson (b) and baryon (c).}
\label{fig:string history}
\end{figure*}

\subsection{The splitting matrix $T$}\label{sec:new T for quarks}
To describe the elementary splitting $\Q\rightarrow h+\Qp$ in momentum and spin-space, we generalize the splitting matrix for the splitting $q\rightarrow h+\qp$ introduced in the string+${}^3P_0$ model of Ref. \cite{Kerbizi:2021M20}. We write the splitting matrix as
\begin{eqnarray}\label{eq:T quark}
\nonumber && T_{\Qp,\h,\Q}(\mh,Z,\pt;\kt) = C_{\Qp,\h,\Q}\,D_{\h}(\mh) \\
\nonumber &\times& f^{1/2}_{\Qp,h,\Q}(Z,\pt;\kt)\, N_{\aQp\aQ}^{-1/2}(\eh^2)\,
\fT(\kptkpt) \\
&\times& \Delta_{\Qp}(\kpt)\,\Gamma_h\,\uhat_{\Q}^{-1/2}(\kt),
\end{eqnarray}
where $\Delta_{\Qp}$ is the propagator of $\Qp$, $\Gamma_h$ is a generally rectangular matrix connecting the spin state of $\Q$ with that of the compound system $(\Qp,h)$, and $h$ denotes a hadron in a given spin state.

The coefficient $C_{\Qp,\h,\Q}$ describes the splitting in flavor-space and is based on the wave function of $\h$ in isospin space.
The function $|D_{\h}(\mh)|^2$ gives the invariant mass distribution of $\h$. For a fixed hadron mass it is a delta function centered on the nominal squared mass. For a resonance it is a relativistic Breit-Wigner function with mass and width fixed to their nominal values.
The $Z$- and $\pt$- dependent part of the splitting matrix is the square-root of the spinless splitting function of the LM~\cite{Andersson:1983jt}, here indicated by $f_{\Qp,h,\Q}$. The latter is given by
\begin{eqnarray}\label{eq:fq'hq mesons}
\nonumber    f_{\Qp,h,\Q}(Z,\pt;\kt)&=&\left(\frac{1-Z}{Z}\right)^{\aQp}\,\left(\frac{Z}{\eh^2}\right)^{\aQ}\\
&\times& \exp(-\bL\eh^2/Z),
\end{eqnarray}
and it describes the distribution of the longitudinal momentum of $\h$ in the splitting $\Q\rightarrow h + \Qp$. It depends on the free parameters $\aQp$, $\aQ$ and $\bL$.

The function $N_{\aQp\aQ}(\eh^2)$, given by
\begin{eqnarray}\label{eq:N_aDaq}
N_{\aQp,\aQ}(\eh^2)=\int_0^1\,dZ\,Z^{-1}\,f_{\Qp,h,\Q}(Z,\pt;\kt),
\end{eqnarray}
plays the role of a normalization factor for the $Z$-dependent part of the splitting amplitude squared. It depends on the squared transverse energy of $\h$. In the string+${}^3P_0$ model we assume $a_{\qp}=a_{q}\equiv a$, which leads to $N_{\aqp,\aq}\equiv N_a(\eh^2)$ \cite{Kerbizi:2021M20}. For diquarks we take $\aqq=\aqqbar\equiv \aD$ with $\aD\neq a$, as in the implementation of the LM in the Pythia MCEG \cite{Bierlich:2022pfr}.

The function $\fT$ in Eq. (\ref{eq:T quark}) provides the cutoff for the relative transverse momentum of the $\qp\qpbar$ pair created at the string breaking (see also Fig. \ref{fig:tunneling}a) and it is taken to have the exponential form $\fT(\kptkpt)=\sqrt{\bT\,\pi^{-1}}\exp(-\bT\,\kptkpt/2)$, with $\bT$ being a free parameter. $|\fT|^2$ is the generalization of the formula $P_{\rm{Tun.}}(\kptkpt)=\exp\left[-\pi(m_{q'}^2+\kptkpt)/\kappa\right]$ found in Ref. \cite{Casher:1978wy} for the tunneling probability of the quark pair, with $\kappa\simeq 0.2\,\GeV^{2}$ being the string tension or equivalently the energy per unit length stored in the string. For the tunneling of a $(\qq)(\qqbar)$ pair the same expression holds with $m_{q'}\rightarrow m_{\qq}$. We thus assume that the function $\fT$ describes also the transverse momentum cutoff for the diquarks. 

We rewrite the product $\Delta_{\Qp}\,\Gamma_h$ in Eq. (\ref{eq:T quark}) in terms of a $2\times 2$ matrix $t_{\Qp,h,\Q}$ connecting only the two spin-1/2 particles. More precisely, for each type of splitting, we define $t_{\Qp,h,\Q}$ by
\begin{eqnarray}
\nonumber    \langle &&\Qp,S_{\Qp};h,S_h|\,\Delta_{\Qp}\,\Gamma_h|\Q,\SQ\rangle=\\
\nonumber   && \begin{cases}
        \chi^{\dagger}(\Sqp)\,t_{\qp,h,q}(\kpt)\,\chi(\Sq), & \Qp=\qp,\,h=M\\
        \chi^{\dagger}(\SB)\,t_{(\qqbar),B,q}(\kpt)\,\chi(\Sq), & \Qp=\qqbar,\,h=B\\
        \chi^{\dagger}(\Sqp)\,t_{\qp,\Bbar,(\qqbar)}(\kpt)\,\eta(\SBbar), & \Qp=\qp,\,h=\Bbar
    \end{cases}.\\
\end{eqnarray}
We have introduced the Pauli spinors $\chi(\Sq)$, $\chi(\Sqp)$, $\chi(\SB)$ and $\eta(\SBbar)=\sigma_z\,\chi(-\SBbar)$~\footnote{This is the analogue of the relation $v(k,s)=\gamma_5\,u(k,-s)$ involving Dirac spinors.} of, respectively, $q$, $\qp$, $B$ and $\Bbar$. $\chi(\textbf{S})$ indicates the Pauli spinor with polarization vector $\textbf{S}$.

The matrix-function $\uhat_{\Q}$ in Eq. (\ref{eq:T quark}) is given by \cite{Kerbizi:2021M20}
\begin{eqnarray}\label{eq:uhat_general}
\nonumber    \uhat_{\Q}(\kt) &=& \sum_{h,s_h}\,|C_{\Qp,h,\Q}|^2\,\int\, d^2\kpt\,\int\,\frac{dZ}{Z}\,f_{\Qp,h,\Q}(Z,\pt;\kt)\\
    &\times& N^{-1}_{\aQp,\aQ}(\eh^2)\,\fT^2(\ktkt)\,t^{\dagger}_{\Qp,h,\Q}(\kpt)\,t_{\Qp,h,\Q}(\kpt),
\end{eqnarray}
where the summation is taken over the produced hadron $h$ (meson and baryon) and on its spin states. An integration over $\mh^2$ is also understood. It can be shown that $\uhat_{\Q}\propto 1_{2\times 2}$ for $\Q=q$ and $\uhat_{\Q}\propto 1_{3\times 3}$ for $Q=(\qqbar)_1$ due to the factor $N_{\aQp,\aQ}(\eh^2)$ in Eq. (\ref{eq:N_aDaq}). The proportionality factor is a constant factor and it can be obtained once the expression for $\t_{\Qp,h,\Q}$ is given.

The explicit expressions for $t_{\Qp,h,\Q}$ for the splittings in Eq. (\ref{eq:splittings}) are as follows.

\paragraph{Splitting $q\rightarrow M+\qp$.}
The matrix-function $t_{\qp,h,q}$ for $h=\PS,\VM$ can be written as \cite{Kerbizi:2019ubp,Kerbizi:2021M20}
\begin{eqnarray}\label{eq:t quark}
    t_{\qp,h,q}(\kpt) = \Delta_{\qp}(\kpt)\, \Gamma_{h,s_h},
\end{eqnarray}
where the $2\times 2$ matrix $\Delta_{q'}(\kpt)$ is the propagator of the quark $q'$ obtained by the ${}^3P_0$ mechanism, and the $2\times 2$ matrix $\Gamma_{h,s_h}$ indicates the coupling of $q$ and $q'$ with the meson $h$. The explicit expression for $\Delta_{\qp}$ is given in Eq. (\ref{eq:Deltaq}), while the expression for $\Gamma_{h,s_h}$ is given in Eq. (\ref{eq:Gamma q}).

\paragraph{Splitting $q\rightarrow B+\Qp$.} The matrix-function $t_{\Qp,B,q}(\kpt)$ can be written as
\begin{eqnarray}\label{eq:t}
 \nonumber   t_{\Qp,B,q}(\kpt)=\begin{cases}
        \Phi^*_{a}\,\Delta_{\qq,ab}(\kpt)\,\Gamma_{B,b}, & \Qp=(\qqbar)_1 \\
        \Gamma_{B,0}, & \Qp=(\qqbar)_0.
    \end{cases} \\
\end{eqnarray}
 $\boldsymbol{\Phi}$ is the spin wavefunction of the PV antidiquark whose components in the cartesian basis are $\Phi_{a}$, $a=x,y,z$. $\Gamma_{B,b}$, with $b=x,y,z$, and $\Gamma_{B,0}$ are $2\times 2$ matrices that describe the couplings $q-B-(\qq)_1$ and $q-B-(\qq)_0$, respectively. The $3\times 3$ matrix $\Delta_{\qq}(\kpt)$ is the propagator for the PV antidiquark $\qqbar$. The explicit expressions for the couplings $\Gamma$ are given in Eq. (\ref{eq:Gamma baryon}), while the expressions for the propagator $\Delta_{\qqbar}$ is given in Eq. (\ref{eq:Delta_qq final}).

\paragraph{Splitting $(\qqbar)\rightarrow \Bbar + \qp$.} The expression for the matrix $t_{\qp,\Bbar,(\qqbar)}$ is
\begin{eqnarray}\label{eq:t diquark}
\nonumber    t_{\qp,\Bbar,(\qqbar)}(\kpt) =
        \begin{cases}
        \Delta_{\qp}(\kpt)\,\sigma_z\,\Gamma_{B,b}\,\sigma_z\,\Phi_{b}, & \,(\qqbar)=(\qqbar)_1 \\
        \Delta_{\qp}(\kpt)\,\Gamma_{B,0}\, & \,(\qqbar)=(\qqbar)_0
        \end{cases},\\
\end{eqnarray}
where the quark propagator $\Delta_{\qp}$ and the vertex $\Gamma_{B}$ are the same as in Eqs. (\ref{eq:t quark})-(\ref{eq:t}).

\paragraph{Splittings $\qbar\rightarrow \Bbar+(\qq)$ and $(\qq)\rightarrow B +\qbarp$.} Due to the charge conjugation symmetry,
the splitting matrices for the splittings $\qbar\rightarrow \Bbar+(\qq)$ or $(\qq)\rightarrow B+\qbarp$ can be obtained from Eq. (\ref{eq:T quark}) with the substitutions $Q\rightarrow \qbar$, $h\rightarrow\Bbar$ and $\Qp\rightarrow (\qq)$ or $\Q\rightarrow (\qq)$, $h\rightarrow B$ and $\Qp\rightarrow \qbarp$. 
These splittings can be applied to the description of the target remnant fragmentation in a deep-inelastic scattering event, which will be studied in a separate work.

The splitting matrix for $\qbar\rightarrow M +\qbarp$ can be found in Ref. \cite{Kerbizi:2023luv}.

\subsection{The coupling matrix $\Gamma$}\label{sec:Gamma}
For the splitting $q\rightarrow h+q'$, with $h$ a meson, we write according to Eq. (\ref{eq:t quark}) \cite{Kerbizi:2019ubp,Kerbizi:2021M20}
\begin{eqnarray}
    \langle h, \qp|\Gamma|q\rangle = \chi^{\dagger}(\Sqp)\,\Gamma_{h,s_h}\,\chi(\Sq),
\end{eqnarray}
with 
\begin{equation}\label{eq:Gamma q}
   \Gammaq_{\h,\sh} =  \begin{cases}
   \sigma_{z}  & \rm{if\, \h = PS }\\ 
    \GL\,\VL^*\,\Iden + \GT\,\VT^*\cdot\sigmaqT\sigma_{z} & \rm{if\, \h = VM}             
   \end{cases}. 
\end{equation}
With $\boldsymbol{\sigma}_{\T}=(\sigma_x,\sigma_y)$ we indicate the vector of Pauli matrices with transverse components $\boldsymbol{\sigma}_{\rm T}$, and $\V =(\VT,\VL)$ is the vector amplitude of the VM wave function. In the case of VM emission, the coupling matrix depends on the complex coupling constants $\GL$ and $\GT$, which describe the coupling of $\q$ and $\qp$ with a VM having longitudinal and transverse polarization with respect to the string axis, respectively. $\V$ is defined in the VM rest frame obtained by the string frame by the sequence of boosts shown in Ref. \cite{Kerbizi:2021M20}.

For the splitting $q\rightarrow B +\Qp$, we write according to the Eq. (\ref{eq:t})
\begin{eqnarray}\label{eq:reduced coupling amplitude B}
\nonumber    \langle B, \Qp|\Gamma|q\rangle = \begin{cases}
        \chi^{\dagger}(\SB)\,\Gamma_{B,0}\,\chi(\Sq), & \rm{\Qp=(\qqbar)_0}\\
        \chi^{\dagger}(\SB)\,\Phi^{*}_{b}\,\Gamma_{B,b}\,\chi(\Sq) &\rm{\Qp=(\qqbar)_1}
    \end{cases},\\
\end{eqnarray}
with
\begin{eqnarray}\label{eq:Gamma baryon}
    \Gamma_{B,b}&=&\begin{cases}
        \sigmabf_b, & b=x,y,z \\
        1_{2\times 2}, & b=0
    \end{cases}.
\end{eqnarray}

The coupling in Eq. (\ref{eq:Gamma baryon}) is obtained by rewriting the covariant amplitude $\bar{u}_{B}\,V_{\qqbar,B,q}\,u_q$ associated to the splitting $q\rightarrow B+\qqbar$, where $V_{\qqbar,B,q}$ is the coupling matrix, in terms of Pauli spinors in the baryon rest frame [Eq. (\ref{eq:reduced coupling amplitude B})]. This frame is reached by the sequence of boosts introduced in Ref. \cite{Kerbizi:2021M20}. 
We have indicated by $u_B$ and $u_q$ the on-shell Dirac spinors of $B$ and of $q$, respectively.
Beforehand the quark spinor $u_q$ has been projected on the subspace with $\alpha_z=-1$, suitable for a quark moving with velocity $v_z=\langle \alpha_z\rangle \simeq -1$ when entering the baryon. This is inspired by the string fragmentation model where the quarks trajectories are composed of segments with $v_z=dz/dt=\pm 1$ (see Fig. \ref{fig:string history}).

For $V_{\qqbar,B,q}$ we have taken $V_{(\qqbar)_0,B,q}=\,1_{4\times 4}$ for a scalar diquark, and $V_{(\qqbar)_1,B,q}=\gamma_5\,\gamma^{\mu}\,\eps^*_{\mu}$ for a PV diquark with polarization four-vector $\eps^{\mu}$ \cite{Bacchetta:2008af}. Then considering the on-shell relation $\varepsilon^0=\textbf{v}\cdot \boldsymbol{\varepsilon}$ and neglecting the transverse velocity $\textbf{v}_{\rm T}$ we take $\varepsilon^0=\varepsilon^z$, \textit{i.e.}, $\varepsilon^{\mu}=(\varepsilon^z,\varepsilon^x,\varepsilon^y,\varepsilon^z)$. Thus the spin-degree of freedom of the antidiquark can be encoded in the 3-vector $\boldsymbol{\Phi}=(\varepsilon_x, \varepsilon_y, 2\varepsilon_z)$~\footnote{We checked that the alternative coupling $V_{(\qqbar)_1,B,q}=\gamma_5(\gamma^{\mu}+p^{\mu}/M_B)\,\eps^*_{\mu}$ introduced in Ref. \cite{Jakob:1997wg} and suitable for a non-relativistic diquark in the baryon rest frame gives the same coupling in Eq. (\ref{eq:Gamma baryon}) with $\epsbf_{z} = \eps^z$. We also obtained the same result using the wave functions of the spin-1/2 baryons in spin $\otimes$ isospin space in the non-relativistic quark model.}.
This is sufficient to implement in the model the required symmetries, shown in Sec. \ref{sec:symmetries}.

Analogously we obtain the coupling for the splitting $\qqbar\rightarrow \Bbar + q'$, with $Q$ being an antidiquark $(\qqbar)$. In this case we write
\begin{eqnarray}
\nonumber    \langle \Bbar, \qp|\Gamma|Q\rangle = \begin{cases}
        \chi^{\dagger}(\Sqp)\,\Gamma_{B,0}\,\eta(\SBbar), & \rm{\Q=(\qqbar)_0}\\
        \chi^{\dagger}(\Sqp)\,\Phi_{b}\,\sigma_z\Gamma_{B,b}\sigma_z\,\eta(\SBbar) &\rm{\Q=(\qqbar)_1}
    \end{cases},\\
\end{eqnarray}
where the matrix $\Gamma_B$ is the same as in Eq. (\ref{eq:Gamma baryon}).

\subsection{The propagator $\Delta$}\label{sec:propagator}
\begin{figure}[b]
\centering
\begin{minipage}[b]{0.49\textwidth}
\includegraphics[width=0.7\textwidth]{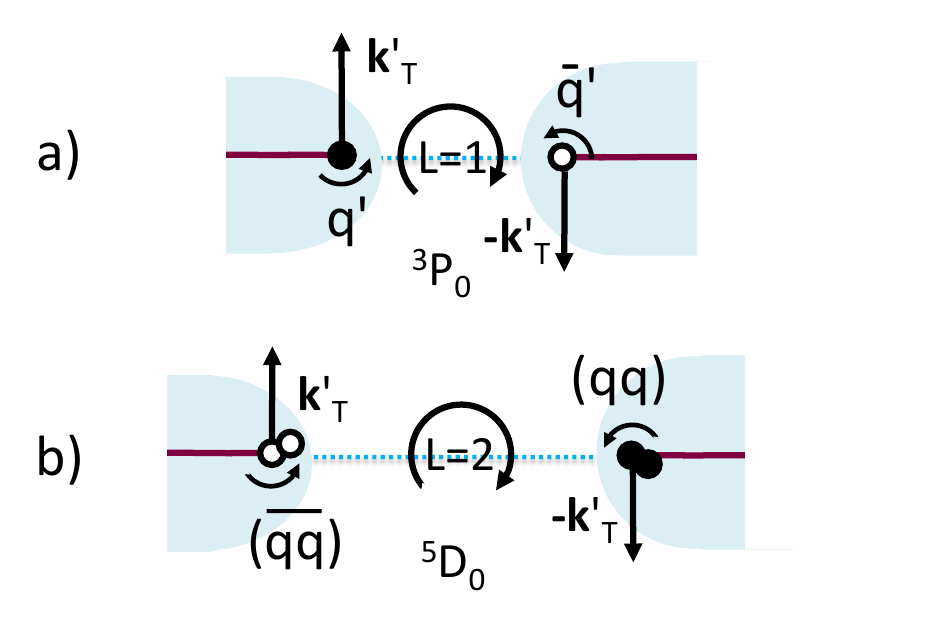}
\end{minipage}
\caption{Quark-antiquark pair tunneling in the ${}^3P_0$ state (a), and diquark-antidiquark pair tunneling in the ${}^5D_0$ state (b). Straight arrows indicate transverse momenta, while the curved arrows indicate either the quark/diquark spin or orbital angular momentum. The dotted line indicates the tunneling length.}
\label{fig:tunneling}
\end{figure}

The quark propagator in Eq. (\ref{eq:t quark}) is \cite{Kerbizi:2021M20}
\begin{eqnarray}\label{eq:Deltaq}
    \Delta(\kpt)=\mu_q+\sigmaZq\boldsymbol{\sigma}\cdot\kpt,
\end{eqnarray}
which parametrizes the ${}^3P_0$ wave function of the $\qp\qpbar$ pair produced at the string breaking (see, \textit{e.g.}, Fig. \ref{fig:tunneling}a). It is given in terms of the complex parameter $\mu_q$. This parameter replaces $k'_z=-\i\,\sqrt{m_{q'}^2+\kptkpt}$ of the quark in the middle of the tunneling trajectory. $\mu_q$ can in principle depend on the flavor of $\qp$ and on $\kptkpt$, but we take it to be flavour-independent and constant. The imaginary part $\rm{Im}(\mu_q)$ is responsible for transverse spin effects, \textit{e.g.} the Collins effect and the dihadron production asymmetry, while $\rm{Im}(\mu_q^2)=2\,\rm{Re}(\mu_q)\,\rm{Im}(\mu_q)$ is responsible for longitudinal spin effects, \textit{e.g.} the jet-handedness \cite{Kerbizi:2018qpp}.

To construct the diquark propagator in Eq. (\ref{eq:t}) we assume that string breakings can occur by tunneling of diquark-antidiquark pair $(\qq)(\qqbar)$ from the string medium, as shown in Fig. \ref{fig:tunneling}b, with vacuum quantum numbers $J^{PC}=0^{++}$. $J=0$ implies $L=S$, where $L$ is the relative orbital angular momentum of the pair and $\textbf{S}=\textbf{S}(\qq)+\textbf{S}(\qqbar)$ is the total spin. The parity of the $(\qq)-(\qqbar)$ pair is $P=(-1)^L$ and its charge conjugation $C=(-)^{L+S}$. Therefore $L=S=0$ (${}^1S_0$ state) or $L=S=2$ (${}^5D_0$ state). As a matter of fact the string medium is not isotropic; there is a chromoelectric field that can produce a kind of Stark effect, \textit{i.e.}, ad admixture of $J=2,4,\dots$ states. We ignore such an effect.

Starting from a pure ${}^5D_0$ state of the diquark pair and using the fact that in the middle of the tunneling trajectory we have $E^2_{\qq}=m_{\qq}^2+\kv'^2=0$ we arrive at a propagator of the form (see Appendix \ref{App:diquark propagator})
\begin{eqnarray}\label{eq:Delta_qq final}
 \nonumber   \Delta_{\qq}(\kpt) = \begin{pmatrix}
        \kpxkpx+\mqq^2/3 & \kpx\kpy & \muqq\,\kpx \\
        \kpx\kpy & \kpykpy+\mqq^2/3 & \muqq\,\kpy \\
        \muqq\,\kpx & \muqq\,\kpy & \muqq^2+\mqq^2/3
    \end{pmatrix}, \\
\end{eqnarray}
where we have replaced first $\kv'^2$ by $-m_{\qq}^2$, then the remaining $k'_z$ by a phenomenological complex parameter $\muqq$. The propagator $\Delta_{\qq}$ is thus a $3\times 3$ matrix that depends on the transverse momentum $\kpt$ of the diquarks at the exit of the tunneling and on the parameter $\muqq$. The latter is the analogue of the complex parameter $\mu_q$ introduced for the tunneling of quarks via the ${}^3P_0$ mechanism [see Eq. (\ref{eq:Deltaq})]. The real and imaginary parts of $\muqq$ are free parameters of the model whose value must be fixed by comparison with data.

\section{Probability distribution for the produced hadron and the Collins effect}\label{sec:probability distribution and collins effect}
\subsection{The splitting functions of the model}
The splitting matrix in Eq. (\ref{eq:T quark}) is related to the splitting probability by
\begin{eqnarray}\label{eq:dP general}
    \frac{dP_{\Q\rightarrow h+\Qp}}{d\mh\,dZZ^{-1}\,d^2\pt} &=& |\langle \Qp,S_{\Qp};h,\Sh| T_{\Qp,h,\Q} |\Q,\SQ \rangle|^2\\
\nonumber    &\equiv& F_{\Qp,h,\Q}(Z,\pt;\kt,\rho(\Q),\check{\rho}(\Qp),\check{\rho}(h)),
\end{eqnarray}
where we have indicated by $|\Q,\SQ\rangle$, $|h,\Sh\rangle$ and $|\Qp,S_{\Qp}\rangle$ the spin states of $\Q$, $h$ and $\Qp$, respectively. 
The function $F_{\Qp,h,\Q}(Z,\pt;\kt,\rho(\Q),\rhocheck(\Qp),\rhocheck(h))$ is the triple-polarized splitting function and it describes the energy-momentum sharing between the emitted hadron $h$ and the leftover particle $\Qp$, when all involved particles are polarized. With $\rho(X)$ we indicate the spin density matrix of a given particle $X=q,\qqbar$, e.g. $\rho(q)=(\Iden+\sigmabf\cdot\Sq)/2$ for a quark $q$ with polarization vector $\Sq$. The symbol $\rhocheck$ indicates the acceptance spin density matrix (for the definition and use of such matrix see, \textit{e.g.}, Ref. \cite{Artru:2008cp}). $\rhocheck(h)$ implements the information on the directions of the decay products of $h$ or on their accepted angular domain. $\rhocheck(\Qp)$ implements the spin information coming "backwards in time" from the successive emissions of $\Qp$. The latter information is neglected in the recursive string+${}^3P_0$ model, namely we take $\rhocheck(\Qp)=1$. It is used, at least temporarily, as long as one has no information about its future decay or splitting. For $\rhocheck(h)=1$, the resulting function $F_{\Qp,h,\Q}(Z,\pt;\kt,\rho(\Q))$ is referred to as the (\textit{polarized}) splitting function. It gives the probability distribution for emitting $h$ from a polarized $\Q$.

\paragraph{Case $q\rightarrow M+\qp$.} Inserting Eq. (\ref{eq:t quark}) in Eq. (\ref{eq:T quark}), and using Eq. (\ref{eq:dP general}) with $h=\PS,\VM$ yields the triple-polarized splitting function
\begin{eqnarray}\label{eq:splitting function M}
\nonumber   && F_{\qp,h,q}(Z,\pt;\kt,\rho(q),\rhocheck(\qp),\rhocheck(h)) = \\
\nonumber &\times& |C_{\qp,h,q}|^2\,|D_h(\mh)|^2\, f_{\qp,h,q}(Z,\pt;\kt)\,N^{-1}_{a}(\eh^2)\\
 &\times& \uhat_q^{-1}\,\Tr_{\qp}\left[\Delta_{\qp}\,\Gamma^a_{h,s_h}\,\rho(q)\,\Gamma^{\dagger\,b}_{h,s_h}\,\Delta^{\dagger}_{\qp}\,\check{\rho}(\qp)\right] \,\check{\rho}_{ba}(h),
\end{eqnarray}
The trace is taken over the spin indices of $q'$, and we have indicated with $\check{\rho}(\qp)$ the acceptance density matrix of $\qp$. For a PS meson emission $\rhocheck_{ba}(h)$ and the indices $a,b$ are removed. For VM emission we write $\Gamma_{h,s_h}=\Gamma^a_{h,s_h}\,V^*_{a}$, with $a=x,y,z$ [see Eq. (\ref{eq:Gamma q})].

\paragraph{Case $q\rightarrow B+(\qqbar)$.} The triple-polarized splitting function for baryon production via the splitting $q\rightarrow B+\qqbar$ can be evaluated inserting Eq. (\ref{eq:T quark}) and Eq. (\ref{eq:t}) for $h=\B$ in Eq. (\ref{eq:dP general}). We obtain
\begin{eqnarray}\label{eq:splitting function B}
\nonumber    && F_{\qqbar,B,q}(Z,\pt;\kt,\rho(q),\rhocheck(B),\rhocheck(\qqbar)) =  |D_{B}(m_B^2)|^2\\
\nonumber &\times& |C_{\qqbar,B,q}|^2\, f_{\qqbar,B,q}(Z,\pt;\kt)\,N^{-1}_{\aDbar,a}(\varepsilon_B^2)\\
\nonumber    &\times& \uhat_q^{-1}\,\Delta_{\qq;\rm ab}(\kpt)\,\Tr_{B}\left[\Gamma_{B,b}\,\rho(q)\,\Gamma_{B,b'}^{\dagger}\,\check{\rho}(B)\right]\,\Delta_{\qq;b'a'}^{\dagger}(\kpt),\\
&\times& \check{\rho}_{a'a}(\qqbar)
\end{eqnarray}
where $\check{\rho}(B)$ is the acceptance matrix of $B$ and $\check{\rho}(\qqbar)$ is the acceptance matrix of $(\qqbar)$ (it is a $3\times 3$ matrix for a PV diquark and $1$ for a scalar diquark). The trace is taken over the Pauli spin indices of $B$.

\paragraph{Case $(\qqbar)\rightarrow +\Bbar+\qp$.}
For the antidiquark splitting $(\qqbar)\rightarrow \Bbar+\qp$, the triple-polarized splitting function is obtained similarly from Eq. (\ref{eq:dP general}). We obtain
\begin{eqnarray}\label{eq:splitting function qq->B}
\nonumber    && F_{\qp,\Bbar,(\qqbar)}(Z,\pt;\kt,\rho(\qqbar),\rhocheck(\qp),\rhocheck(\Bbar)) =  |D_{\Bbar}(\MBbar)|^2 \\
\nonumber &\times& |C_{\qp,\Bbar,(\qqbar)}|^2\,f_{\qp,\Bbar,(\qqbar)}(Z,\pt;\kt)\,N^{-1}_{a,\aD}(\varepsilon_{\Bbar}^2)\\
\nonumber    &\times& \uhat_{(\qqbar)}^{-1}\,\Tr_{\qp}\bigg[\Delta_{\qp}(\kpt)\,\sigma_z\Gamma_{B,a}\sigma_z\,\check{\rho}(\Bbar)\,\sigma_z\Gamma_{B,b}^{\dagger}\sigma_z\,\Delta^{\dagger}_{\qp}(\kpt)\\
&\times& \check{\rho}(\qp)\bigg]\,
 \rho_{ab}(\qqbar),
\end{eqnarray}
where the trace is taken over the leftover quark spin indices. $\rho_{ab}(\qqbar)$ is the spin density matrix of the fragmenting PV antidiquark. $\rhocheck(\Bbar)=\sigma_z\rho(-
\SBbarcheck)\sigma_z$ is the acceptance matrix of $\Bbar$, with $\SBbarcheck$ the corresponding acceptance polarization vector. For a scalar diquark we remove the $\sigma_z$ matrices, $\rho_{ab}(\qqbar)$, and take $a=b=0$.


\begin{figure}[tbh]
\centering
\includegraphics[width=0.5\textwidth]{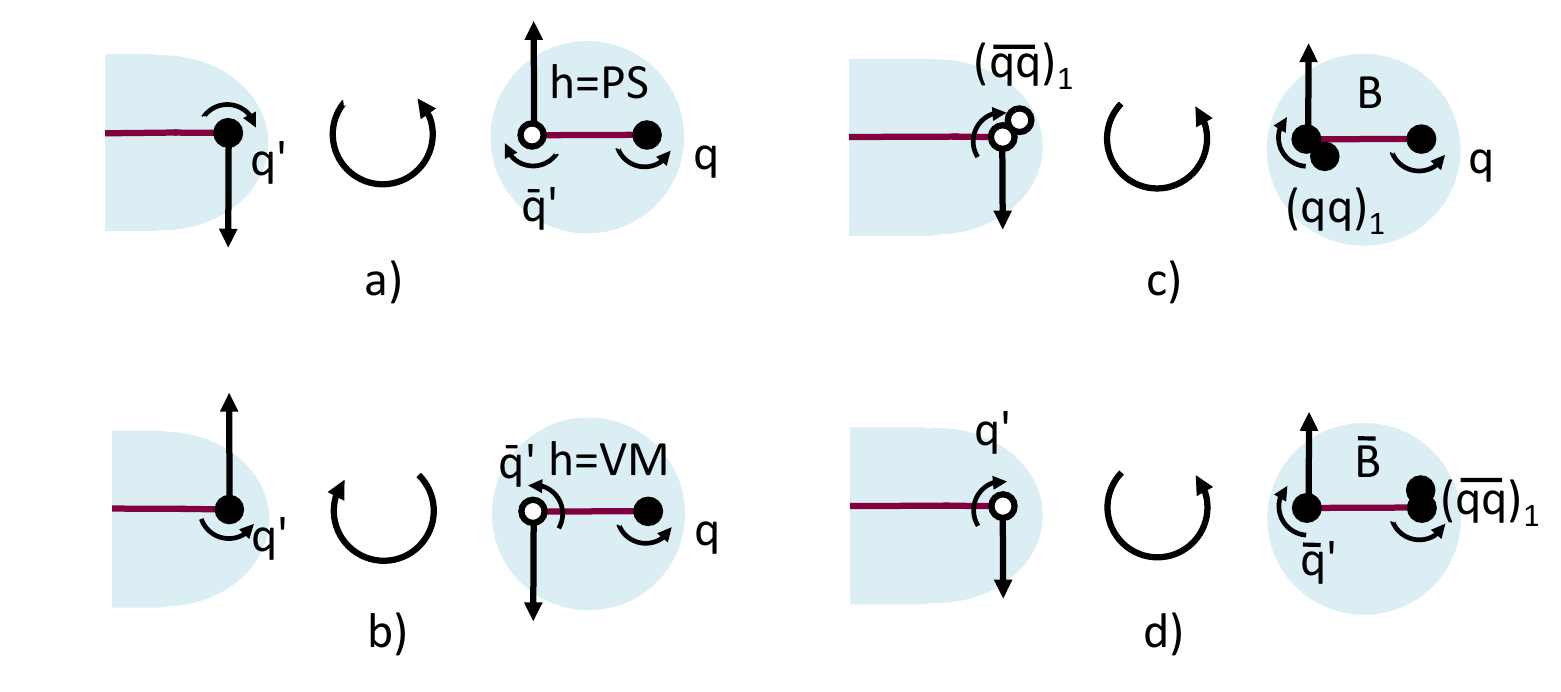}
\caption{Fragmentation of a string with a transversely polarized initial quark (a, b, c) or antidiquark (d). String+${}^3P_0$ mechanism for PS meson (a), VM (b) and antibaryon (d) production. String+${}^5D_0$ mechanism for baryon production (c).}
\label{fig:string+5D0}
\end{figure}

\subsection{Explicit splitting function for meson production}
The splitting function associated to the quark splitting $\q\rightarrow h+\qp$ with $h$ being a PS meson or a VM has been extensively studied in Ref. \cite{Kerbizi:2021M20}. We recall it in this section, to simplify the comparisons with the new splittings $q\rightarrow B +(\qqbar)$ and $(\qq)\rightarrow B +\qbarp$.

As a preliminary step the matrix $\uhat_q$ in Eq. (\ref{eq:uhat_general}) with $\Qp=q'$ can be calculated by inserting Eqs. (\ref{eq:Deltaq})-(\ref{eq:Gamma q}) in Eq. (\ref{eq:t quark}), and by using the obtained expression of $t_{q',h,q}$ in Eq. (\ref{eq:uhat_general}). We decompose the result as \cite{Kerbizi:2021M20} 
\begin{align}\label{eq:u matrix q meson}
\nonumber   \hat{u}_{\q} =& \sum_h\displaystyle\,\hat{u}_{\q,h}\,1_{2\times 2}, \\
\nonumber   \hat{u}_{\q,\h} =& |C_{\qp,\h,\q}|^2\,(|\mu_q|^2+\langle \kTkT\rangle_{\fT})\, \\
   & \times \begin{cases}
   1  & \rm{if\, h = PS }\\ 
   \fVM & \rm{if\, h = VM}                 
   \end{cases}, 
\end{align}
where for a generic function $H(\ktkt)$ we have defined $\langle H(\kTkT)\rangle_{\fT} = \int\,d^2\kT\,H(\kTkT)\,\fT^2(\kTkT)$. The constant $\fVM/(1+\fVM)$, with $\fVM=|2|\GT|^2+|\GL|^2$, is the probability for the $\q\qbarp$ pair to be a VM rather than a PS meson. It is one of the free parameters of the LM implemented in Pythia \cite{Bierlich:2022pfr}.

The splitting function can be obtained inserting Eq. (\ref{eq:u matrix q meson}) in Eq. (\ref{eq:splitting function M}), and taking $\check{\rho}(q')=1_{2\times 2}$ and $\check{\rho}(h)=1_{3\times 3}$ for a VM or $\rhocheck(h)=1$ for a PS meson. It reads \cite{Kerbizi:2021M20}
\begin{align}\label{eq:splitting F_VM explicit}
\nonumber    F&_{\qp,h,\q}(\mh,Z,\pt;\kt,\SqT)
= \frac{\hat{u}_{\q,h}}{\hat{u}_{\q}}\, |D_{\h}(\mh)|^2\, \\
\nonumber &\times \left(\frac{1-Z}{\eh^2}\right)^{a}\,e^{-\bL\eh^2/Z}\, N_a^{-1}(\eh^2)\, \\
\nonumber &\times \fT^2(\kptkpt)\,\frac{|\mu_q|^2+\kptkpt}{|\mu_q|^2+\langle \kTkT\rangle_{\fT}}\\
\nonumber &\times \left[1+\,c\,\frac{2\Im(\mu_q)}{|\mu_q|^2+\kptkpt}\,\SqT\cdot\left(\zu\times \kpt\right)\right].\\
\end{align}
The first line describes the relative probability of producing the meson $\h$, given by the ratio $\hat{u}_{\q,\h}/\hat{u}_{\q}$, and the mass distribution of $h$. The second line gives the distribution of the longitudinal splitting variable $Z$. The distributions of $\kptabs$ and the azimuthal angle of $\kpt$ mainly result, respectively, from the third and last lines (they are also affected by the $\eps_h^2$-dependence). Thus $\kpt$ has a mean orientation which is perpendicular to the transverse polarization $\SqT$ of the fragmenting quark. The amplitude of this effect depends on the parameter $c$, with $c=-1$ for $h=\PS$ and $c=\fL$ for $h=\VM$. The parameter $\fL=|\GL|^/(2|\GT|^2+|\GL|^2)$ describes the fraction of VMs with longitudinal polarization with respect to the string axis.

For the initial fragmenting quark we have $\kt=0$, so we can replace $\kpt$ by $-\pt$ in Eq. (\ref{eq:splitting F_VM explicit}). Comparing with Eq.~(\ref{eq:Collins FF}) we see that the string+${}^3P_0$ model predicts a Collins effect of amplitude
\begin{equation}\label{eq:aM}
   \hat{a}_{C,h}(\ptabs) = -2\,c\,\Im(\mu_q)\,\frac{\ptabs}{|\mu_q|^2+\ptabsptabs},
\end{equation}
which, for a produced PS meson ($c=-1$), is of the experimentally observed sign. Note that the Collins effect for VMs is of opposite sign and with reduced magnitude with respect to the PS meson case \cite{Kerbizi:2021M20}.

\subsection{Explicit splitting function for baryon production}
\subsubsection{Splitting $q\rightarrow B+(\qqbar)_1$}
For the production of a baryon $B$ in the splitting $q\rightarrow B + (\qqbar)_1$, where $(\qqbar)_{1}$ indicates a PV anti-diquark, the matrix $\uhat^{(1)}_{q}$ can be obtained by first inserting Eq. (\ref{eq:Gamma baryon}) and Eq. (\ref{eq:Delta_qq final}) in the expression for $t_{(\qqbar),B,q}$ in Eq. (\ref{eq:t}), and then using the obtained $t_{(\qqbar),B,q}$ in Eq. (\ref{eq:uhat_general}). We write the resulting matrix $\uhat^{(1)}_{q}$ as
\begin{align}\label{eq:u matrix q baryon}
\nonumber   \hat{u}^{(1)}_{\q} =& \sum_B\displaystyle\,\hat{u}^{(1)}_{\q,B}\,1_{2\times 2}, \\
\hat{u}^{(1)}_{\q,\B} =& |C_{(\qqbar)_1,B,\q}|^2\, \langle H_1^2(\ktkt)+ H_2(\ktkt)\rangle_{\fT},                
\end{align}
where we have defined the functions
\begin{eqnarray}
 \nonumber   \Hone(\ktkt) &=& |\muqq|^2+\ktkt, \\
 \Htwo(\ktkt) &=& \frac{\mqq^2}{3}\left(\mqq^2 + 2\ktkt + 2\Re(\muqq^2)\right).
\end{eqnarray}

The splitting function for $q\rightarrow B+(\qqbar)_1$ can be obtained inserting in Eq. (\ref{eq:splitting function B}) the $q-B-(\qqbar)_1$ coupling in Eq. (\ref{eq:Gamma baryon}) and the diquark propagator in Eq. (\ref{eq:Delta_qq final}). Taking $\check{\rho}(B)=1_{2\times 2}$ and $\check{\rho}(\qqbar)=1_{3\times 3}$ we obtain
\begin{align}\label{eq:splitting F q->B+qq1}
\nonumber    &F_{(\qqbar)_1,B,\q}(\MB,Z,\pt;\kt,\SqT)
= \frac{\hat{u}^{(1)}_{\q,B}}{\hat{u}_{\q}}\, |D_{\B}(\MB)|^2\, \\
\nonumber &\times \left(\frac{1-Z}{Z}\right)^{\aD}\,\left(\frac{Z}{\epsB^2}\right)^a\,e^{-\bL\epsB^2/Z}\, N_{\aD,a}^{-1}(\epsB^2)\, \\
\nonumber &\times \fT^2(\kptkpt)\,\frac{\Hone^2(\kptkpt) + \Htwo(\kptkpt)}{\langle \Hone^2(\ktkt) + \Htwo(\ktkt)\rangle_{\fT}}\\
\nonumber &\times \left[1+\,\frac{2\Im(\muqq)\,\Hone(\kptkpt)\,\,\SqT\cdot\left(\zu\times \kpt\right)}{\Hone^2(\kptkpt)+\Htwo(\kptkpt)}\right].\\
\end{align}
The first line of the splitting function describes the splitting in flavour space and the invariant mass distribution of $B$. The second line gives the longitudinal momentum distribution of $B$, which depends on the parameter $a$, $\aD$ and  $\bL$. This is at variance with the splitting function for meson production in Eq. (\ref{eq:splitting function M}), which depends only on the parameters $a$ and $\bL$.

The distribution of the modulus $\kptabs$ of the transverse momentum of $(\qqbar)_1$ is mainly given by the third line of Eq. (\ref{eq:splitting F q->B+qq1}) (it is also affected by the $\eps_B^2$-dependence). It is a fourth order polynomial in $\kptabs$ multiplied by the exponential function $\fT^2(\kptkpt)$, and depends also on the diquark mass $\mqq$, which enters the diagonal elements of the PV diquark propagator in Eq. (\ref{eq:Delta_qq final}). This differs with respect to the meson emission case in Eq. (\ref{eq:splitting F_VM explicit}), which involves a second order polynomial in $\kptabs$.

The last line of the splitting function in Eq. (\ref{eq:splitting F q->B+qq1}) includes the mixed product $\SqT\cdot(\zu\times \kpt)$ responsible for a Collins effect for the emission of the baryon. 
For the initial fragmenting quark we replace $\kpt$ by $-\pt$ in Eq. (\ref{eq:splitting F q->B+qq1}). Comparing with Eq.~(\ref{eq:Collins FF}), the predicted Collins effect has the amplitude
\begin{eqnarray}\label{eq:aB}
\hat{a}^{(1)}_{B}(\ptabs) &=& -2\,\Im(\muqq)\,\frac{\ptabs\,\Hone(\ptabsptabs)}{\Hone^2(\ptabsptabs) + \Htwo(\ptabsptabs)},
\end{eqnarray}
which depends on $\Im(\muqq)$. Assuming $\Im(\muqq)<0$, the Collins effect for the baryon production is predicted to have the same sign as that expected from the classical string+${}^5D_0$ model and shown in Fig. \ref{fig:string+5D0}. In addition, it is predicted to have the same sign as the Collins effect for PS meson emission in the splitting $q\rightarrow h+q'$.

For a typical diquark mass $\mqq\sim 0.5\,\GeV$, the contribution of the $\mqq$-dependent terms in the denominator in Eq. (\ref{eq:aB}) is expected to be small. In the limit of vanishing diquark mass, Eq. (\ref{eq:aB}) reduces to $\hat{a}^{(1)}_B\simeq 2\Im(\muqq)\,\kptabs/[|\muqq|^2+\kptkpt]$. In this limit the Collins analysing power for baryon production in the splitting $q\rightarrow B + (\qqbar)_1$ has the same form as the analysing power for meson production in the splitting $q\rightarrow h + q'$ shown in Eq. (\ref{eq:aM}), provided that the substitution $\mu_q\rightarrow \muqq$ is performed. This can be understood in the non-relativistic quark model of baryons where the diquark plays the role of a color anti-triplet, like the antiquark in the wave function of a meson.

\subsubsection{Splitting $q\rightarrow B+(\qqbar)_0$}
For the splitting $q\rightarrow B+(\qqbar)_0$ we start by evaluating the matrix $\uhat^{(0)}_q$ by inserting Eq. (\ref{eq:t}) in Eq. (\ref{eq:uhat_general}) and using the coupling in Eq. (\ref{eq:Gamma baryon}) for $b=0$. We obtain
\begin{align}\label{eq:u matrix q baryon qq0}
\hat{u}^{(0)}_{\q} = \sum_B\displaystyle\,\hat{u}^{(0)}_{\q,B}\,1_{2\times 2},&& \hat{u}^{(0)}_{\q,\B} = |C_{(\qqbar)_1,B,\q}|^2.
\end{align}

The splitting function can be obtained from Eq. (\ref{eq:splitting function B}) by taking $\check{\rho}(B)=1_{2\times 2}$, removing the diquark propagator and the acceptance matrix $\check{\rho}(\qqbar)$, and using the coupling for scalar diquarks in Eq. (\ref{eq:Gamma baryon}). We obtain
\begin{align}\label{eq:splitting F q->B+qq0}
\nonumber    &F_{(\qqbar)_0,B,\q}(\MB,Z,\pt;\kt,\SqT)
= \frac{\hat{u}^{(0)}_{\q,B}}{\hat{u}_{\q}}\, |D_{\B}(\MB)|^2\, \\
 &\times \left(\frac{1-Z}{Z}\right)^{\aD}\,\left(\frac{Z}{\epsB^2}\right)^a\,e^{-\bL\epsB^2/Z}\, N_{\aD,a}^{-1}(\epsB^2)\, \fT^2(\kptkpt).
\end{align}
As can be seen, for a baryon produced in the splitting $q\rightarrow B +(\qqbar)_0$ the new model predicts a distribution for $\kpt$ much simpler than for a baryon produced in the splitting $q\rightarrow B + (\qqbar)_1$ given in Eq. (\ref{eq:splitting F q->B+qq1}). In particular, for the scalar diquark case the distribution of $\kptabs$ is mainly given by $\fT^2(\kptkpt)$ and the distribution of the azimuthal angle of $\kpt$ is flat. No Collins effect is thus predicted when the baryon is emitted via the tunneling of scalar diquarks.

\paragraph{The effective baryon spectrum.}
The effective baryon spectrum of the model is obtained by summing the splitting functions in Eq. (\ref{eq:splitting F q->B+qq1}) and in Eq. (\ref{eq:splitting F q->B+qq0}), hence
\begin{eqnarray}\label{eq:F total q->B+qqbar}
    F_{(\qqbar),B,q}=F_{(\qqbar)_1,B,q} + F_{(\qqbar)_0,B,q}.
\end{eqnarray}
In the summation each diquark species enters with the weight given by the flavour wave function of $B$, which is encoded in the coefficients $C_{\qqbar,B,q}$.
In particular the Collins analysing power in Eq. (\ref{eq:aB}) is diluted by the scalar diquark contribution.

\subsection{Explicit splitting function for antibaryon production}
\subsubsection{Splitting $(\qqbar)_1\rightarrow \Bbar + \qp$}
The $\uhat_{(\qqbar)}$ matrix for the PV diquark splitting $(\qqbar)_1\rightarrow \Bbar+\qp$ can be evaluated inserting Eq. (\ref{eq:t diquark}) in Eq. (\ref{eq:uhat_general}). We obtain
\begin{align}\label{eq:u matrix (qq)1 baryon}
\nonumber   \hat{u}_{(\qqbar)_1} =& \sum_{\Bbar}\displaystyle\,\hat{u}_{(\qqbar)_1,\Bbar}\,1_{3\times 3}, \\
   \hat{u}_{(\qqbar)_1,\Bbar} =& \,3\,|C_{\qp,\Bbar,(\qqbar)_1}|^2\, \langle |\mu_q|^2+ \kTkT\rangle_{\fT}.                
\end{align}

Inserting Eq. (\ref{eq:u matrix (qq)1 baryon}) and Eq. (\ref{eq:t diquark}) in the splitting function (\ref{eq:splitting function qq->B}), and taking $\check{\rho}(\Bbar)=1_{2\times 2}$ and $\check{\rho}(\qp)=1_{2\times 2}$ we obtain the splitting function
\begin{align}\label{eq:splitting F qq1->B+qbar}
\nonumber    &F_{\qp,\Bbar,(\qqbar)_1}(\MBbar,Z,\pt;\kt,\SqqbarT)
= \frac{\hat{u}_{(\qqbar)_1,\Bbar}}{\hat{u}_{(\qqbar)_1}}\, |D_{\Bbar}(\MBbar)|^2\, \\
\nonumber &\times \left(\frac{1-Z}{Z}\right)^{a}\,\left(\frac{Z}{\epsBbar^2}\right)^{\aD}\,e^{-\bL\epsBbar^2/Z}\, N_{a,\aD}^{-1}(\epsBbar^2)\, \\
\nonumber &\times \fT^2(\kptkpt)\,\frac{|\mu_q|^2+\kptkpt}{\langle |\mu_q|^2+\ktkt\rangle_{\fT}}\\
&\times \left[1-\,\frac{2\Im(\mu_q)}{|\mu_q|^2+\kptkpt}\,\SqqbarT\cdot\left(\zu\times \kpt\right)\right].
\end{align}
The vector $\SqqbarT$ is the transverse component of the vector polarization $\textbf{S}_{\qqbar}$ of the antidiquark with respect to the string axis. The components of $\textbf{S}_{\qqbar}$ are obtained from the spin density matrix of $\qqbar$ as 
\begin{equation}\label{eq:Sqq}
S_{\qqbar,c}=\i\,\varepsilon_{abc}\,\rho_{ab}(\qqbar),
\end{equation}
with $a,b,c=x,y,z$.

Comparing Eq. (\ref{eq:splitting F qq1->B+qbar}) and Eq. (\ref{eq:splitting F_VM explicit}) with $c=-1$ it can be seen that the distribution of the transverse momentum $\kpt$ of the leftover quark $\qp$ in the splitting $(\qqbar)_1\rightarrow \Bbar+\qp$ is the same as that of the leftover quark in the splitting $q\rightarrow \PS + \qp$ where a PS meson is produced. This is in agreement with the classical string+${}^5D_0$ mechanism in Fig. \ref{fig:string+5D0}d.

\subsubsection{Splitting $(\qqbar)_0\rightarrow \Bbar + \qp$}
Inserting Eq. (\ref{eq:t diquark}) in Eq. (\ref{eq:splitting function qq->B}) the splitting function for a scalar antidiquark reads
\begin{align}\label{eq:splitting F qq0->B+qbar}
\nonumber    &F_{\qp,\Bbar,(\qqbar)_0}(\MBbar,Z,\pt;\kt)
= \frac{\hat{u}_{(\qqbar)_0,\Bbar}}{\hat{u}_{(\qqbar)_0}}\, |D_{\Bbar}(\MBbar)|^2\, \\
\nonumber &\times \left(\frac{1-Z}{Z}\right)^{a}\,\left(\frac{Z}{\epsBbar^2}\right)^{\aD}\,e^{-\bL\epsBbar^2/Z}\, N_{a,\aD}^{-1}(\epsBbar^2)\, \\
&\times \fT^2(\kptkpt)\,\frac{|\mu_q|^2+\kptkpt}{\langle |\mu_q|^2+\ktkt\rangle_{\fT}},
\end{align}
where
\begin{align}\label{eq:u matrix (qq)0 baryon}
\nonumber   \hat{u}_{(\qqbar)_0} =& \sum_{\Bbar}\displaystyle\,\hat{u}_{(\qqbar)_0,\Bbar}, \\
   \hat{u}_{(\qqbar)_0,\Bbar} =& |C_{\qp,\Bbar,(\qqbar)_0}|^2\, \langle |\mu_q|^2+\ktkt\rangle_{\fT}.           
\end{align}
$\uhat_{(\qqbar)_0}$ is obtained by inserting Eq. (\ref{eq:t diquark}) with $(\qqbar)=(\qqbar)_0$ in Eq. (\ref{eq:uhat_general}).

For $\kt=\textbf{0}$, Eq. (\ref{eq:splitting F qq0->B+qbar}) results in a flat distribution for the azimuthal angle of $\kpt$, as expected by the fact that the fragmenting scalar antidiquark does not carry any spin information. The distribution of the magnitude of the transverse momentum $\kpt$ is similar to that of the leftover quark in the elementary splitting $q\rightarrow h +\qp$ where the meson $h$ is produced, since in both amplitudes enters the quark propagator in Eq. (\ref{eq:Deltaq}).

\begin{figure}[tbh]
\centering
\includegraphics[width=0.4\textwidth]{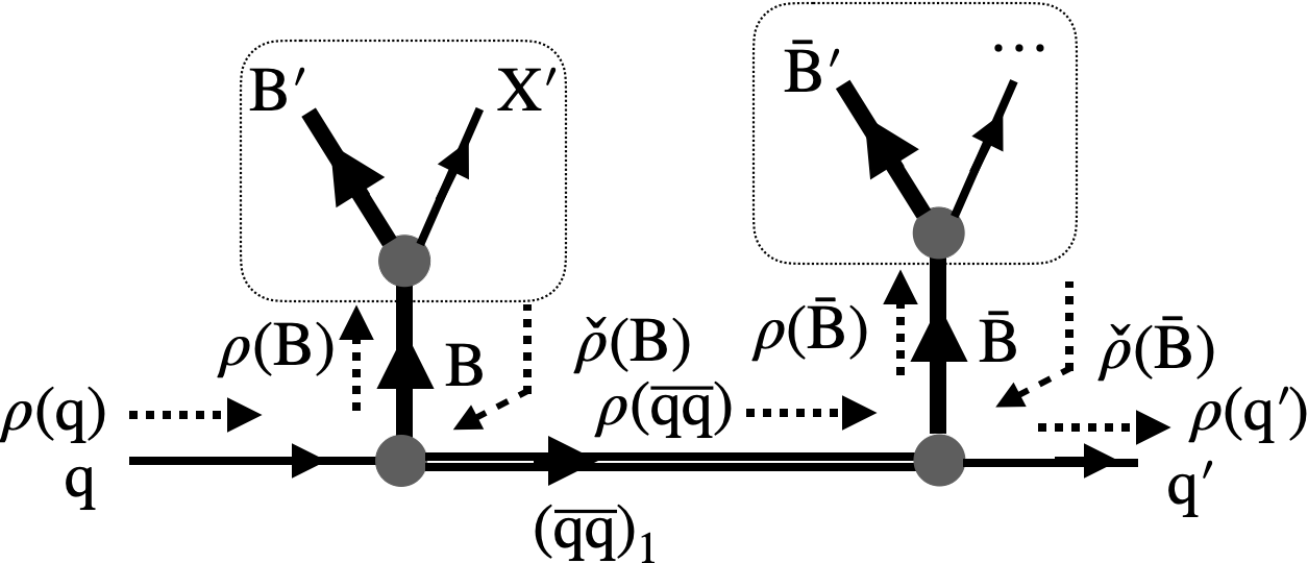}
\caption{Production of a baryon $B$ and antibaryon $\bar{B}$ in the fragmentation chain initiated by the quark $q$. The dotted arrows show the propagation of the spin information by means of spin density matrices and acceptance matrices. The decaying baryons are taken to be hyperons.}
\label{fig:chain}
\end{figure}

\section{Baryon polarization}\label{sec:polarization}
\subsection{The spin density matrix of the baryon}
The spin density matrix of the baryon $B$ produced in the quark splitting $q\rightarrow B+(\qqbar)_1$, shown in the fragmentation chain in Fig. \ref{fig:chain}, can be obtained from the last line of the all-polarized splitting function in Eq. (\ref{eq:splitting function B}), with $\uhat_q=\uhat_q^{(1)}$ of Eq. (\ref{eq:u matrix q baryon}), by summing over the polarization states of the PV diquark (\textit{i.e.}, $\check{\rho}(\qqbar)=1_{3\times 3}$) and interpreting the trace operation as $\rm{Tr}[\rho^{(1)}(B)\,\check{\rho}(B)]$. The baryon spin density matrix $\rho^{(1)}(B)$ is then given by
\begin{eqnarray}\label{eq:rho(B) in q->B+qq1}
   \rho^{(1)}(B) = \frac{\Delta_{ab}\,\Gamma^b\,\rho(q)\,\Gamma^{b'\,\dagger}\,\Delta^{\dagger}_{b'a}}{\Tr[\dots]},
\end{eqnarray}
where the dots in the denominator represent the same expression as in the numerator, and the trace is taken over the baryon spin indices.

By inserting in Eq. (\ref{eq:rho(B) in q->B+qq1}) the explicit expression for the diquark propagator in Eq. (\ref{eq:Delta_qq final}) and coupling in Eq. (\ref{eq:Gamma baryon}), the spin density matrix $\rho^{(1)}(B)$ of the emitted baryon can be explicitly evaluated. The polarization vector $\SB^{(1)}$ of the baryon can be evaluated as $\SB^{(1)}=\Tr[\boldsymbol{\sigma}\,\rho^{(1)}(B)]$. We use the decomposition $\SB^{(1)}=(\SbTone,\SbLone)$ with $\SbTone$ and $\SbLone$ being the transverse and longitudinal components with respect to the string axis, respectively.

We obtain for the transverse component of the baryon polarization
\begin{eqnarray}\label{eq:SBT axial diquark}
\nonumber   &&\SbTone =  \frac{1}{\it{N}_{\q}(\Sq)}\,\bigg[-2\Im(\muqq)\,\Hone(\kptkpt)\,\zu\times \kpt \\
\nonumber&-& \SqT\,\left(\Hone^2(\kptkpt) + \Htwo(\kptkpt)\right)\\ 
\nonumber &+& 2\,(\kpt\cdot\SqT)\kpt\,\bigg(\Hone(\kptkpt)+\frac{2\mqq^2}{3}\bigg) \\
   &+& 2\Re(\muqq)\,\Sqz\,\kpt\,\bigg(\Hone(\kptkpt)+\frac{2\mqq^2}{3}\bigg)\bigg],
\end{eqnarray}
where the normalization function $N_{\q}(\Sq)$ is given by
\begin{eqnarray}
 \nonumber   N_{\q}(\Sq) &=& \Hone^2(\kptkpt) +\Htwo(\kptkpt)\\
 &+& 2\Im(\muqq)\,\Hone(\kptkpt)\,\SqT\cdot (\zu\times \kpt).
\end{eqnarray}

For a baryon emitted by the initial quark we have $\kpt=-\pt$ and the first term in Eq. (\ref{eq:SBT axial diquark}) gives a baryon transverse polarization in the direction $\zu\times \pt$. Taking an unpolarized $q$ and comparing with Eq. (\ref{eq:polarizing FF}), the spontaneous polarization of $B$ results
\begin{eqnarray}\label{eq:a_B spont q split}
    \hat{a}^{(1)}_{B}(\ptabs) = 2\,\Im(\muqq)\,\frac{\ptabs\,\Hone(\ptabsptabs)}{\Hone^2(\ptabsptabs)+\Htwo(\ptabsptabs)}
\end{eqnarray}
For $\Im(\muqq)<0$ the baryon spontaneous is negative, in agreement with the BELLE data on the spontaneous polarization of $\Lambda$ and $\bar{\Lambda}$ hyperons produced in $e^+e^-$ annihilation. Note that the size of the spontaneous polarization  in Eq. (\ref{eq:a_B spont q split}) depends on $\Im(\muqq)$, which could be determined by a more quantitative comparison with data. Such comparison would require the implementation of the model in a MC event generator, which will be presented in a separate work.

The terms in the second line of Eq. (\ref{eq:SBT axial diquark}) give the transfer of transverse polarization from the quark to the baryon. In particular the baryon acquires a transverse polarization which is opposite to that of the quark. This contribution is in competition with that given by the propagation of scalar diquarks in Eq. (\ref{eq:SBT q->B+Dbar0}). A study of the transverse polarization transfer from quarks to $\Lambda$ and $\bar{\Lambda}$ hyperons was performed by the COMPASS experiment using SIDIS data with a transversely polarized proton target \cite{COMPASS:2021bws}.
The last term in Eq. (\ref{eq:SBT axial diquark}) gives instead a conversion of the longitudinal polarization of the quark to the transverse polarization of the baryon.

For the longitudinal polarization $(S_{B,\rm{L}}\equiv S_{B,z})$ of the baryon we find
\begin{eqnarray}\label{eq:SB scalar diquark}
\nonumber    &&\SbLone = \bigg\{ -\left(\textbf{k}^{'4}_{\rm T}-|\muqq|^4+\frac{\mqq^2}{3}\left(\frac{\mqq^2}{3}+2\kptkpt\right)\right)\,\Sqz \\
 \nonumber &+& 2\Re(\muqq)\,\SqT\cdot\kpt\,\bigg(\Hone(\kptkpt)+\frac{2\mqq^2}{3}\bigg)\bigg\}\times N_q^{-1}(\Sq).\\
\end{eqnarray}
The first line gives the transfer of longitudinal polarization from the quark to the baryon, while the second line describes the conversion of the quark transverse polarization to the longitudinal polarization of the baryon.

If the leftover diquark in the splitting $q\rightarrow B+(\qqbar)_0$ is a scalar diquark then, according to Eqs. (\ref{eq:splitting function B}) and (\ref{eq:Gamma baryon}), the spin density matrix of the baryon is 
\begin{equation}\label{eq:rho(B) q->B+qq0}
\rho^{(0)}(B)=\rho(q),
\end{equation}
which leads to the baryon polarization vector
\begin{eqnarray}\label{eq:SBT q->B+Dbar0}
    \SbTzero = \SqT, \,\,\,\,\,\,\,\,\,\,\,\,\,\,\, \SbLzero = \SqL.
\end{eqnarray}
Hence the quark polarization is completely transferred to the baryon, due to the fact that the scalar diquark does not carry spin information.

Comparing with Eqs. (\ref{eq:SBT axial diquark})-(\ref{eq:SB scalar diquark}), one can see that the contribution to the transverse spin transfer from the quark to the baryon via a leftover scalar diquark has opposite sign with respect to the case of a leftover PV diquark.

\paragraph{The effective spin density matrix of the baryon.} The effective spin density matrix for the baryon $B$ produced in the splitting $q\rightarrow B +(\qqbar)$ predicted by the model is obtained by the weighted sum of Eq. (\ref{eq:rho(B) in q->B+qq1}) and Eq. (\ref{eq:rho(B) q->B+qq0}). We obtain
\begin{eqnarray}
    \rho(B)=\frac{F_{(\qqbar)_1,B,q}\,\rho^{(1)}(B) + F_{(\qqbar)_0,B,q}\,\rho^{(0)}(B)}{F_{(\qqbar)_1,B,q} + F_{(\qqbar)_0,B,q}}.
\end{eqnarray}
The polarization vector of $B$ can then be obtained analogously by the substitutions $\rho^{(1)}(B)\rightarrow\SB^{(1)}$ and $\rho^{(0)}(B)\rightarrow \SB^{(0)}$.

\subsection{The spin density matrix of the antibaryon}
The spin density matrix of the antibaryon $\Bbar$ produced in the splitting $(\qqbar)_1\rightarrow \Bbar +\qp$ can be obtained from the triple-polarized splitting function in Eq. (\ref{eq:splitting function qq->B}) by taking $\check{\rho}(\qp)=1_{2\times 2}$ and interpreting the spin dependent part as $\Tr[\rho^{(1)}(\Bbar)\check{\rho}(\Bbar)]$, with $\rho^{(1)}(\Bbar)=\sigma_z\rho^{(1)}(-\SBbar)\sigma_z$. The baryon spin density matrix is thus
\begin{eqnarray}\label{eq:rho(B) in qq1 splitting}
   \rho^{(1)}(\Bbar) &=& \frac{\sigma_z\,\Gamma^{\dagger}_{B,b}\,\sigma_z\,\Delta^{\dagger}_{\qbar'}\,\Delta_{\qbar'}\,\sigma_z\Gamma_{B,a}\,\sigma_z\,\rho_{ab}(\qq)}{\Tr[\dots]},
\end{eqnarray}
where the trace operation in the denominator is taken over the baryon spin indices.
The resulting transverse and longitudinal components of the baryon polarization can be derived by using the quark propagator in Eq. (\ref{eq:Deltaq}) and the coupling in Eq. (\ref{eq:Gamma baryon}). The transverse component of the polarization vector reads
\begin{eqnarray}\label{eq:SBT in qq1 splitting}
\nonumber    \SBbarT^{(1)} &=& \frac{1}{N_{\qqbar}(\Sqqbar)}\,\bigg[(|\mu_q|^2+\kptkpt)\,\SqqbarT-2\Im(\mu_q)\,\zu\times\kpt \\
&+& 4\Im(\mu_q)\, \left(\Re\rho_{\rm T}(\qqbar)\,[\zu\times\kpt]\right)\bigg],
\end{eqnarray}
where $\rho_{\rm T}(\qqbar)$ is the $3\times 3$ matrix with components $\rho_{\rm T, ij}(\qqbar)=\rho_{ij}(\qqbar)$ for $i,j=x,y$ and zero elsewhere. The vector $\Sqq$ is defined in Eq. (\ref{eq:Sqq}). The normalization function is given by
\begin{eqnarray}\label{eq:N(qq)}
   N_{\qqbar}(\Sqqbar) = |\mu_q|^2+\kptkpt-2\Im(\mu_q)\,\SqqbarT\cdot (\zu\times \kpt).\,\,\,\,\,
\end{eqnarray}

The first term in Eq. (\ref{eq:SBT in qq1 splitting}) describes the transfer from diquark transverse polarization to the baryon transverse polarization. The second term is a source of spontaneous polarization for the baryon [c.f. with Eq. (\ref{eq:polarizing FF})] and, since in the string+${}^3P_0$ model $\Im(\mu_q)>0$, the polarization has the same sign as that arising in the splitting $q\rightarrow B + (\qqbar)_1$ in Eq. (\ref{eq:SBT axial diquark}). The model thus predicts the same sign for the spontaneous polarization of baryons and antibaryons, as observed for $\Lambda$ and $\bar{\Lambda}$ hyperons in $e^+e^-$ annihilation at BELLE \cite{Belle:2018ttu}. The last term describes the transfer of the transverse tensor polarization from the diquark to the baryon. 

The baryon longitudinal polarization resulting from Eq. (\ref{eq:rho(B) in qq1 splitting}) is
\begin{eqnarray}\label{eq:SBL in qq1 splitting}
\nonumber \SBbarL^{(1)} &=& \frac{1}{N_{\qqbar}(\Sqqbar)}\,\bigg[(|\mu_q|^2+\kptkpt)\,\SqqbarL\\
&+&4\,\Im(\mu_q)\,\zu\cdot \,\left(\Re\rho(\qqbar)\,[\zu\times\kpt]\right)\bigg],
\end{eqnarray}
where $\SqqbarL=\i\,[\rho_{xy}(\qqbar)-\rho_{yx}(\qqbar)]$ is the longitudinal vector polarization of the diquark. The first term describes the transfer of longitudinal polarization from the diquark to the baryon while the second term proportional to $\Re(\rho_{zy})\,k'_x-\Re(\rho_{zx})\,k'_y$ describes the conversion of the diquark oblique polarization to the baryon longitudinal polarization.

Concerning the scalar antidiquark splitting $(\qqbar)_0\rightarrow \Bbar +\qp$, the corresponding spin density matrix of the produced baryon can be obtained from Eq. (\ref{eq:rho(B) in qq1 splitting}) by removing $\rho_{ab}(\qqbar)$ and taking for $\Gamma_{B,0}$ the unit matrix according to Eq. (\ref{eq:Gamma baryon}). The baryon spin density matrix reads
\begin{eqnarray}\label{eq:rho(B) qq0}
\nonumber   \rho^{(0)}(\Bbar) &=& \frac{\Delta^{\dagger}_{\qbar'}\,\Delta_{\qbar'}}{\Tr[\dots]}= \frac{1}{2}\,\left[1 + \frac{2\Im(\mu_q)}{|\mu_q|^2 + \kptkpt}\,\sigmaT\cdot(\zu\times \kpt)\right],\\
\end{eqnarray}
which gives for the polarization vector of the antibaryon
\begin{eqnarray}\label{eq:SBT scalar diquark}
    \SBbarT^{(0)} = \frac{2\Im(\mu_q)}{|\mu_q|^2+\kptkpt}\, \zu\times \kpt, \,\,\,\,\,\,\,\,\,\,\,\, \SBbarL^{(0)} = 0.
\end{eqnarray}
Baryons produced in scalar antidiquark splittings have therefore spontaneous polarization, which originates from the correlation between the spins and transverse momenta of the quark pair produced via the ${}^3P_0$ mechanism.
Our result in Eq. (\ref{eq:SBT scalar diquark}), obtained with the quantum-mechanical string+${}^3P_0$ model, is reminiscent of the expression for the spontaneous polarization of $\Lambda$ hyperons produced in unpolarized $pp$ scattering obtained by the Lund group using a semiclassical model of hadronization involving the ${}^3P_0$ mechanism \cite{Andersson:1979wj}.

\section{Polarized hyperon decay}\label{sec:hyperon decays}
\begin{figure}[tb]
\centering
\begin{minipage}[b]{0.49\textwidth}
\includegraphics[width=0.45\textwidth]{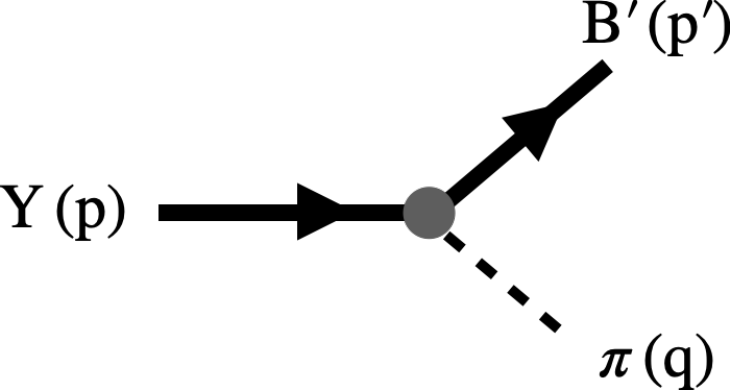}
\end{minipage}
\caption{Diagram for the non-leptonic decay $Y\rightarrow B'+\pi$ of a hyperon $Y$. In parentheses is indicated the four-momentum of each particle.}
\label{fig:decays}
\end{figure}

The spin density matrix $\rho(B)$ [$\rho(\Bbar)$] of the produced baryon (antibaryon) is used for the description of the decay of the particle in its rest frame (see Fig. \ref{fig:chain}). In this work we consider the two-body decays of hyperons $B=Y$, of the type $Y\rightarrow B'+X$, where $B'=p,n,Y'$ is another baryon and $X=\gamma,\pi$ the remaining decay product (see Fig. \ref{fig:decays}). This applies to the hyperons $Y=\Lambda,\Sigma,\Xi$. We distinguish among three types of decays: the non-leptonic (NL) decays ($\Lambda\rightarrow p +\pi$, $\Sigma^+\rightarrow p + \pi^0$, etc.), the electromagnetic (EM) decay $\Sigma^0\rightarrow \Lambda+\gamma$, and the weak radiative (WR) decays ($\Xi^0\rightarrow \Sigma^0+\gamma$, $\Lambda\rightarrow n + \gamma$, etc.). In this section we focus on the NL decay, the most common among the hyperons, and describe the other decays in Appendix \ref{app:decays}.

In the string rest frame, where the four-momentum of $Y$ is $p$, we indicate by $p'$ the four-momentum of $B'$. The decay is described in the rest frame of $Y$. In this frame we indicate the four-momenta of the decay products by capital letters, \textit{i.e.} $P'$ and $Q'$. We define the rest frame of $Y$ by the combination of boosts introduced in Ref. \cite{Kerbizi:2021M20}:
\begin{itemize}\setlength\itemsep{0em}
    \item[i)] a longitudinal boost $B^{-1}_{\rm L}(p_{z}/\epsB)$ that brings $Y$ at a reference frame with $p_{\rm L}=0$, but conserving $\pt$, and
    \item[ii)] a transverse boost $B^{-1}_{\rm T}(\pt/M_Y)$ that brings $Y$ at rest.
\end{itemize}


In the rest frame of $Y$, the baryon $B'$ has four-momentum $P'=(E_{B'},\Pp)$, with $\Pp\equiv|\Pp|\,\boldsymbol{\hat{P}}'$. The two-body kinematics is solved by $E_{B'}=[M^2_Y-M^2_{X}+M^2_{B'}]/(2M_Y)$ and $|\Pp|=\sqrt{E_{B'}^2-M_{B'}^2}$ \cite{ParticleDataGroup:2022pth}. The direction of $B'$ is given by $\boldsymbol{\hat{P}}'=(\cos\phipp\,\sin\thetapp, \sin\phipp\,\sin\thetapp,\cos\thetapp)$, where $\phipp$ and $\thetapp$ are the azimuthal and polar angles of $B'$ in the rest frame of $Y$.

After $\Pphat$ has been generated according to the polarized angular distribution $dN_{Y\rightarrow B'+X}/d\Omega_{\Pp}$ of the decay $Y\rightarrow B'+X$, where $d\Omega_{\Pphat}=d\phipp\,d\cos\thetapp$, $p'$ is obtained by the sequence of boosts $p'=B_{\rm L}(p_z/\epsB)\,B_{\rm T}(\pt/M_B)\,P'$.

\subsection{Non-leptonic decay}\label{sec:NL decays}
To obtain the polarized angular distribution in the rest frame of $Y$, the amplitude associated to the decay $Y\rightarrow B'+X$ is needed. The latter reads
\begin{eqnarray}\label{eq:M reduced}
    \bar{u}_{B'}(p')\,\Mcov^{Y\rightarrow B'+X'}\,u_Y(p) = \chi^{\dagger}(\SBp)\,\Mhat\,\chi(\SY),
\end{eqnarray}
where $\bar{u}_{B'}$ is the onshell Dirac spinor of $B'$ and $\Mcov$ is the covariant decay matrix element. The right-hand side expression is obtained in the rest frame of $Y$. It is the decay amplitude in Pauli spin space, with $\chi(\SBp)$ being the spinor with polarization vector $\SBp$ that describes the spin state of $B'$ in its rest frame. The reduction of $\Mcov$ in Pauli spin space is the $2\times 2$ matrix $\Mhat$. It depends on the decay process and determines the angular distribution of the decay products in the rest frame of $Y$.

For the weak decay $Y\rightarrow B'+\pi$ (Fig. \ref{fig:decays}b) we parametrize the covariant matrix element as $\Mcov\propto [A_s\,1_{4\times 4}-B_p\gamma_5]$, with $A_s$ and $B_p$ being complex parameters \cite{ParticleDataGroup:2022pth}. The parameter $A_s$ gives a parity violating contribution to the decay amplitude. We obtain for the reduced matrix element
\begin{eqnarray}\label{eq:M NL}
    \Mhat_{NL} \propto A_{s} + \tilde{B}_p\,\sigmabf\cdot \Pphat,
\end{eqnarray}
where we have introduced the parameter $\tilde{B}_p=B_p\,|\Pp|/(E_{B'}+M_{B'})$.

The decay distribution in the rest frame of $Y$ for both $Y$ and $B'$ polarized is
\begin{eqnarray}\label{eq:dN NL}
    \frac{dN^{\NL}_{Y\rightarrow B'+\pi}}{d\Omega_{\Pphat}} \propto\Tr_{B'}\left[\Mhat_{\NL}\,\rho(Y)\,\Mhat^{\dagger}_{\NL}\,\check{\rho}(B')\right].
\end{eqnarray}
The explicit decay distribution can be obtained inserting Eq. (\ref{eq:M NL}) in Eq. (\ref{eq:dN NL}). If the detector of $B'$ does not favour a special polarization $\rhocheck(B')=1_{2\times 2}$ and one gets the known angular distribution
\begin{eqnarray}
    \frac{dN_{Y\rightarrow B'+\pi}}{d\Omega_{\Pphat}}=\frac{1}{4\pi}\,\left(1+\alpha_Y\,\SY\cdot\Pphat\right).
\end{eqnarray}
The parameter $\alpha_Y=2\Re{A_s^*\tilde{B}_p}/(|A_s|^2+|\tilde{B}_p|^2)$ depends on the decaying hyperon. It is measured for different hyperons and can be taken from Ref. \cite{ParticleDataGroup:2022pth}.

The spin density matrix of $B'$ can be deduced from Eq. (\ref{eq:dN NL}). We obtain
\begin{eqnarray}
    \rho(B') = \frac{\Mhat_{\NL}\,\rho(Y)\,\Mhat^{\dagger}_{\NL}}{\Tr[\dots]}=\frac{1}{2}\,\big[1+\sigmabf\cdot \SBp\big],
\end{eqnarray}
where the explicit expression for the polarization vector $\SBp$ can be obtained using the matrix element in Eq. (\ref{eq:M NL}) and reads
\begin{eqnarray}\label{eq:SY for NL decay}
\nonumber    \SBp &=& \left(1+\alpha_Y\,\SY\cdot\Pphat\right)^{-1} \,\bigg[(\alpha_Y+\SY\cdot\Pphat)\,\Pphat \\
    &+&\beta_Y\,\SY\times \Pphat + \gamma_Y\,\Pphat\times(\SY\times \Pphat)\bigg].
\end{eqnarray}
The parameters $\beta_Y=2\Im(A_s^*\,\tilde{B}_p)/(|A_s|^2+|\tilde{B}_p|^2)$ and $\gamma_Y=(|A_s|^2-|\tilde{B}_p|^2)/(|A_s|^2+|\tilde{B}_p|^2)$ depend on the decaying hyperon and are related to $\alpha_Y$ by $\alpha_Y^2+\beta_Y^2+\gamma_Y^2=1$. They have been measured for different hyperons and can be found in Ref. \cite{ParticleDataGroup:2022pth}.

Finally, the acceptance matrix of $B$, used in the Collins-Knowles recipe \cite{Collins:1987cp,Knowles:1988vs}, reads
\begin{eqnarray}
\check{\rho}(B) &=& \Mhat^{\dagger}_{\NL}\,\check{\rho}(B')\Mhat_{\NL} \propto 1+\sigmabf\cdot\SYcheck,
\end{eqnarray}
which is evaluated at the generated value of $\Pphat$. The expression for the acceptance polarization vector $\SYcheck$ can be obtained from Eq. (\ref{eq:SY for NL decay}) by the replacements $\SY\rightarrow \SBpcheck$ and $\beta_Y\rightarrow -\beta_Y$. For a not-analysed baryon $B'$ the vector $\SYcheck$ is non-vanishing due to the parity violation in the NL decay, and it reads $\SYcheck=\alpha_Y\,\Pphat$.

\section{Propagation of the spin information along the fragmentation chain}\label{sec:propagation of spin information}
After a baryon $B$ has been emitted in the quark splitting $q\rightarrow B+(\qqbar)$ and $B$ has decayed, it is necessary to propagate the spin information along the fragmentation chain. The propagation of the spin information is shown in Fig. \ref{fig:chain} by the dotted lines ending with an arrow for the chain of splittings $q\rightarrow B +(\qqbar)_1$, $(\qqbar)_1\rightarrow \Bbar+q'$. Such lines represent either spin density matrices or acceptance matrices.

To propagate the spin information along the string fragmentation chain, it is necessary to evaluate the spin density matrix of the leftover particle after each elementary splitting. For the meson emission case $q\rightarrow M +\qp$ this has been done in Ref. \cite{Kerbizi:2021M20}. Here we study the splittings $q\rightarrow B + (\qqbar)_1$ followed by $(\qqbar)_{0,1}\rightarrow \Bbar+\qp$ involving the production of a baryon $B=p,n,Y$.
If the two splittings in Fig. \ref{fig:chain} are connected by a scalar antidiquark $(\qqbar)_0$, the polarization of $q$ is not transferred to $q'$. Even in this case the $\qp$ is polarized and its spin density matrix is needed to propagate the spin correlations along the chain starting from the production of $\Bbar$.

\subsection{Spin density matrix of $(\qqbar)$ in reaction $q\rightarrow B+(\qqbar)$}
The spin density matrix $\rho_{aa'}(\qqbar)$ of the PV anti-diquark $(\qqbar)_1$ in the splitting $q\rightarrow B + (\qqbar)_1$ can be read off from the all-polarized splitting function in Eq. (\ref{eq:splitting function B}), by interpreting the last line as $\rho_{aa'}(\qqbar)\,\check{\rho}_{a'a}(\qqbar)$. It gives
\begin{eqnarray}
    \rho_{aa'}(\qqbar)=\frac{\Delta_{ab}\,\Tr_{B}\left[\Gamma_b\,\rho(q)\,\Gamma_{b'}^{\dagger}\,\check{\rho}(B)\right]\,\Delta_{b'a'}^{\dagger}}{\Tr\left[\dots\right]},
\end{eqnarray}
where the trace appearing in the denominator is taken over the anti-diquark spin indices. As can be seen, the spin state of the antidiquark depends on the spin state of the fragmenting quark $q$ via $\rho(q)$ and on the orientation of the decay products in the decay of $B$ via the acceptance matrix $\rhocheck(B)$ for a hyperon $B=Y$, as expected by the Collins-Knowles recipe \cite{Collins:1987cp,Knowles:1988vs}. For a stable baryon ($B=p,n$) we take $\rhocheck(B)=1_{2\times 2}$.

\subsection{Spin density matrix of $\qp$ in reaction $(\qqbar)\rightarrow \Bbar+\qp$}
The spin density matrix $\rho(q')$ of the leftover quark $\qp$ in the splitting $(\qqbar)\rightarrow \Bbar+\qp$ can be obtained from Eq. (\ref{eq:splitting function qq->B}), by interpreting the last line of the all-polarized splitting function as $\Tr\left[\rho(\qp)\,\check{\rho}(\qp)\right]$. It gives
\begin{eqnarray}
\nonumber    \rho(\qp) \propto\begin{cases}
        \Delta_{\qp}\,\check{\rho}(\Bbar)\,\Delta_{\qp}^{\dagger}, & (\qqbar)_0\\
        \Delta_{\qp}\,\sigma_z\Gamma_{B,a}\sigma_z\,\rho_{aa'}(\qqbar)\,\check{\rho}(\Bbar)\,\sigma_z\Gamma^{\dagger}_{B,a'}\,\sigma_z\,\Delta_{\qp}^{\dagger}, & (\qqbar)_1
    \end{cases}.\\
\end{eqnarray}
As can be seen, in the splitting of a scalar antidiquark $(\qqbar)_0$ the quark $\qp$ can also be polarized despite the absence of spin information coming from the diquark. In particular, for a non analyzed antibaryon $\Bbar$, the polarization of $\qp$ arises from the correlation between spin and transverse momentum at string breaking due to the ${}^3P_0$ mechanism. For a fragmenting PV antidiquark $(\qqbar)_1$, the polarization of $\qp$ depends on $\kpt$, on the spin density matrix $\rho(\qqbar)$ of the antidiquark and on the orientation of the decay products of $\Bbar$, via $\rhocheck(\Bbar)$, for a hyperon $\Bbar=\bar{Y}$ if the decay is analysed.

\subsection{Recursive fragmentation chain}
The complete fragmentation chain is simulated recursively by applying the splittings $q\rightarrow h + q'$, $q\rightarrow B +(\qqbar)$ or $(\qqbar)\rightarrow \bar{B} + q'$ until the energy-momentum in the string falls below some threshold after which the exit condition is called and the fragmentation chain is stopped. The thorough description of the recursive algorithm, including the exit condition, is the same as that provided in Ref. \cite{Kerbizi:2021M20} and is not reported here.

Given a fragmenting quark $q$, two new ingredient are needed to allow for the production of a baryon: the relative probability $P_{\qq}/P_{q}$ for the successive breaking to occur via the tunneling of a $(\qq)-(\qqbar)$ pair rather than the tunneling of a $q'\qbar'$ pair, and the relative probability $P_{(\qq)_1}/P_{(\qq)_0}$ for the diquark to be PV rather than scalar (apart from the enhancement by a factor of three due to the number of states of spin-1 diquarks). These parameters are used to randomly decide whether to emit a baryon by the splitting $q\rightarrow B + (\qqbar)$ or a meson by the splitting $q\rightarrow h + q'$. They are among the free parameters included in the event generator Pythia \cite{Bierlich:2022pfr}, and can be different for non-strange and diquarks involving one or two strange quarks. Typically, $P_{qq}/P_q\sim 0.1$ and $P_{(\qqbar)_1}/P_{(\qqbar)_0}\sim 0.03$. Hence a string breaking occurs via the tunneling of quark-antiquark pairs roughly ten times more frequently than via the tunneling of diquark-antidiquark pairs. Also, PV diquarks at string breakings are further suppressed compared to scalar diquarks. However, for a given baryon, the contribution of PV diquarks can be enhanced by its flavour wavefunction.

\section{Conclusions}\label{sec:conclusions}
We have extended the string+${}^3P_0$ model of spin-dependent string fragmentation by introducing the production and the decay of polarized spin 1/2 baryons. To include baryon production we assume that a quark of the recursive fragmentation chain can be replaced by an antidiquark (or an antiquark by a diquark) of spin $0$ (scalar) or $1$ (pseudovector). The $3\times 3$ propagator of a pseudovector diquark is inspired from a tunneling mechanism producing a diquark-antidiquark pair in the ${}^5D_0$ state while breaking the string, in analogy with the ${}^3P_0$ mechanism of $q\qbar$ pair creation assumed for the quark propagator. Its adopted precise form depends on a complex parameter $\muqq$, which plays the analogue role of the complex mass $\mu_q$ of the ${}^3P_0$ mechanism.
Using the pseudovector diquark propagator and introducing the couplings of quarks and diquarks to baryons, we have constructed the \textit{matrix splitting amplitudes} for the description of the elementary quark and diquark splittings to baryons. They constitute the essential ingredients required for the systematic propagation of the spin correlations along the fragmentation chain.

Using these matrices we show that a Collins effect is predicted for the emission of baryons in the fragmentation of transversely polarized quarks due to the correlation between spin and transverse momentum of the tunneling spin-$1$ diquarks. For $\Im(\muqq)<0$ the Collins effect for baryon production has the the same sign as for the PS meson emission in the string+${}^3P_0$ model, as also predicted by a classical string+${}^5D_0$ mechanism.

Furthermore, we show that the model reproduces a spontaneous transverse polarization for the emitted baryon along the vector perpendicular to the production plane of the baryon, as well as the transverse spin-transfer from a fragmenting transversely polarized quark to the emitted baryon. For a hyperon, the predicted spontaneous polarization has the same sign as that measured in $e^+e^-$ annihilation to hadrons by the BELLE experiment. This is an encouraging result that motivates studies of more quantitative predictions of the model. This calls for the implementation of the model in the Monte Carlo event generator Pythia, which will be presented in a separate work.

Finally, the model presented in this work can be regarded as the basis for new interesting applications such as the description of the spin effects in the target fragmentation region in deep-inelastic scattering events, as well the simulation of the spin-dependent \textit{fracture functions}. We also note that the model allows for the successive production of $B\bar{B}$ pairs, while configurations such as $BM_1M_2\dots\bar{B}$, where mesons $M_1M_2\dots$ are produced between the baryons like in the \textit{pop-corn model} \cite{Casher:1978wy,Andersson:1984af}, are neglected. The production of such final state hadronic configurations can be achieved by either introducing the splittings of (anti)diquarks to mesons and a leftover (anti)diquark, or by endowing the pop-corn model with the spin degree of freedom. The exploration of these applications is planned for future work.

\begin{acknowledgments}
AK is grateful to L. Lönnblad and G. Gustafson for the many interesting and useful discussions on the subject of this work. The work of AK is done in the context of the project “SPINFRAG: Spin-dependent string fragmentation”, funded by the European Union under Marie Skłodowska-Curie Actions (MSCA), grant agreement ID 101107452.
\end{acknowledgments}

\begin{appendix}
\section{Construction of the spin-1 diquark propagator}\label{App:diquark propagator}
\subsection{Theoretical constraints on the propagator}\label{sec:symmetries}
Indicating by $\boldsymbol{\Phi}$ and $\boldsymbol{\Phi}'$ the initial and final (with respect to the propagation) three-dimensional spin wave-functions of the diquark, we require the amplitude $\boldsymbol{\Phi}^{'\dag}\Delta(\kt)\boldsymbol{\Phi}$ to be invariant under
\begin{itemize}\setlength\itemsep{0em}
\item[1)] simultaneous rotation of $\boldsymbol{\Phi}$, $\boldsymbol{\Phi}'$ and $\kt$ about the $\zu$ axis, 
\item[2)] symmetry about any plane containing  the $\zu$ axis,
\item[3)] longitudinal boost of $k$ and $k'$ for fixed $\boldsymbol{\Phi}$ and $\boldsymbol{\Phi}'$,
\item[4)] quark line reversal (LR) symmetry. 
\end{itemize}
The symmetry 2) implies 1). The LR symmetry in 4) is also called \textit{left-right symmetry} in the Lund Model of string fragmentation \cite{Andersson:1983jt}. We find that the LR symmetry implies for the propagator the condition
\begin{eqnarray}
\Delta_{\qqbar}(-R\kv) = R\,\Delta_{\qqbar}^{\rm T}(\kv)\,R,    
\end{eqnarray}
where $R=\rm{diag}(-1,1,-1)$ is the rotation matrix of an angle $\pi$ about the $\yu$ axis.

We find that the required invariances 1)-4) are satisfied by the general form 
\begin{eqnarray}\label{eq:delta decomposition in g_i}
\nonumber    \Delta(\kt) &=& g_1(\ktkt)\,\Delta_1(\kt) + g_2(\ktkt)\,\Delta_2(\kt) \\
&+& g_3(\ktkt)\,\Delta_3(\kt) + g_4(\ktkt)\,\Delta_4(\kt),
\end{eqnarray}
with
\begin{align}\label{eq:D matrices}
\nonumber
&&\Delta_1 = \begin{pmatrix}
1 &  0 & 0
\\ 0 & 1 & 0 
\\ 0 & 0 & 0 
\end{pmatrix}, 
&&
\Delta_2 = \begin{pmatrix}
0 &  0 & 0
\\ 0 & 0 & 0 
\\ 0 & 0 & 1 
\end{pmatrix},\\
&&\Delta_3 = \begin{pmatrix}
\kx^2 &  \kx \ky   & 0
\\ \ky  \kx  & \ky^2   & 0 
\\ 0 & 0 & 0 
\end{pmatrix}, && 
\Delta_4 = \begin{pmatrix}
0 &  0 & \kx 
\\ 0 & 0 & \ky   
\\ \kx  & \ky   & 0 
\end{pmatrix}.
\end{align}

The functions $g_i(\ktkt)$ are a priori unknown, complex and independent. In the following subsection we employ a mechanism for the production of a diquark-antidiquark pair at the string breakings that leads to simple expressions for these functions.


\subsection{Diquark pair production from vacuum: the ${}^5D_0$ mechanism}

Let us start with the two possible wave functions of the $(\qq)_1-(\qqbar)_1$ pair having the $J^{PC}=0^{++}$ quantum numbers of the vacuum
\begin{eqnarray}
    \Psi^{{}^1S_0}_{ab}&\propto& \delta_{ab}\\
    \Psi^{{}^5D_0}_{ab}&\propto& -\frac{\textbf{k}^2}{3}\,\delta_{ab}+\textbf{k}_{a}\textbf{k}_{b},
\end{eqnarray}
where $\R=\kbf(\qqbar)=-\kbf(\qq)$ is the relative momentum of the pair.

The most general $J^{PC}=0^{++}$ wave function is of the form
\begin{eqnarray}
    \Psi_{ab}(\R)&=&A(\R^2)\,\Psi^{{}^1S_0}_{ab}+ B(\R^2)\,\Psi^{{}^5D_0}_{ab}.
\end{eqnarray}

As a temporary choice, let us take for the diquark propagator $\Delta_{qq;\,ab}(\R)=\Psi_{ab}(\R)$, which gives
\begin{eqnarray}\label{eq:delta intermediate}
 &&\Delta_{\qq}(\R) = B(\R^2) \\
\nonumber &\times& \begin{pmatrix}
        \RxRx+D(\R^2) & \Rx\,\Ry & \Rx\,\Rz \\
        \Rx\,\Ry & \RyRy+D(\R^2) & \Ry\,\Rz \\
        \Rz\,\Rx & \Rz\,\Ry & \RzRz+D(\R^2)
    \end{pmatrix},
\end{eqnarray}
with $D(\R^2)=A(\R^2)/B(\R^2)-\R^2/3$. Without loss of generality we take $B(\R^2)=1$, hence $D(\R^2)=A(\R^2)-\R^2/3$. This temporary choice of the propagator does not satisfy Eq. (\ref{eq:delta decomposition in g_i}), because it depends on $\Rz$. For a definitive choice, we must first replace $B(\R^2)$ with $B(\RT^2)$.

The latter function is apriori unknown. A possible choice can be inspired from the semi-classical picture of string breaking in Fig. \ref{fig:tunneling}b. The figure represents the positions on the string and the momenta of $(\qq)$ and $(\qqbar)$ when they exit the tunneling region for a given $\kt$. There, $\Rz=0$ and the orbital angular momentum of the pair is $\textbf{L}=-d\,\zu\,\times\,\kbf$, where $d$ is the $(\qq)-(\qqbar)$ distance. Energy conservation in the tunneling yields $d=2\,\sqrt{\mqq^2+\ktkt}/\kappa$, leading to $\textbf{L}=-2\,\sqrt{\mqq^2+\ktkt}\,\zu\times\kt/\kappa$. Assuming a typical diquark mass $\mqq\sim 0.5\,\GeV$ \cite{Andersson:1984af} and using the value of the string tension $\kappa\simeq 0.2\,\GeV^{2}$, the Schwinger formula for the tunneling probability of the diquark-antidiquark pair (see Sec. \ref{sec:new T for quarks}) yields an average orbital angular momentum $\langle L\rangle\simeq 1.3$. Considering this large average value of $L$, we assume the $(\qq)_1-(\qqbar)_1$ pair to be produced essentially in the ${}^5D_0$ state. This corresponds to taking $A(\kptkpt)=0$ in Eq. (\ref{eq:delta intermediate}).

In the center of the tunneling region, the longitudinal relative diquark momentum is $k_{z}=-\i\,(\mqq^2+\ktkt)^{1/2}$. Replacing first $\R^2$ by $-\mqq^2$, then the remaining $k_z$ by a phenomenological complex parameter $\muqq$, our model result for the spin-1 diquark propagator is
\begin{eqnarray}\label{eq:Delta_qq final appendix}
 \nonumber   \Delta_{\qq}(\kpt) &=& \begin{pmatrix}
        \kpxkpx+\mqq^2/3 & \kpx\kpy & \muqq\,\kpx \\
        \kpx\kpy & \kpykpy+\mqq^2/3 & \muqq\,\kpy \\
        \muqq\,\kpx & \muqq\,\kpy & \muqq^2+\mqq^2/3
    \end{pmatrix}. \\
\end{eqnarray}
Comparing with Eq. (\ref{eq:delta decomposition in g_i}) the model gives $g_1=\mqq^2/3$, $g_2=\muqq^2+\mqq^2/3$, $g_3=1$ and $g_4=\muqq$. Equation (\ref{eq:Delta_qq final appendix}) is thus a particular case of Eq. (\ref{eq:delta decomposition in g_i}). The $L=0$ component can be increased by adding a same constant to $g_1$ and $g_2$. Modifying $g_1-g_2$ introduces an $S=2$ and $J=2$ component. Modifying $g_3$ introduces a $L=1$ component. The last two changes are at variance with the $J=0$ or positive parity assumption and will not be considered.

\begin{figure}[tb]
\centering
\begin{minipage}[b]{0.49\textwidth}
\includegraphics[width=0.45\textwidth]{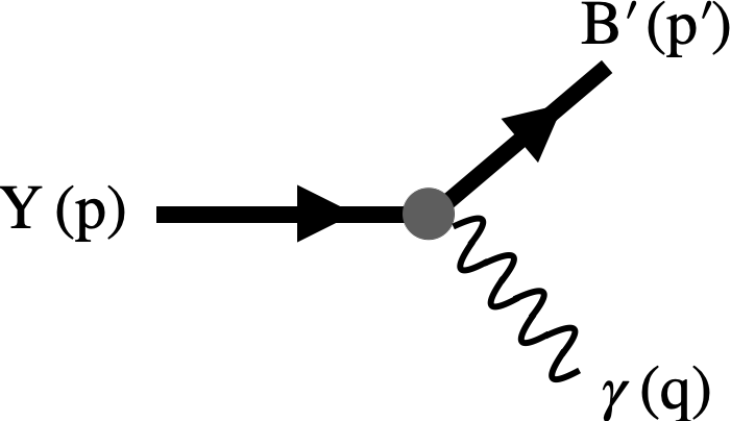}
\end{minipage}
\caption{Diagram for the electromagnetic or radiative weak decay $Y\rightarrow B'+\gamma$ of a hyperon $Y$. In parentheses is indicated the four-momentum of each particle.}
\label{fig:decays EM WR}
\end{figure}

\section{Radiative decays of hyperons}\label{app:decays}
In this section we study the electromagnetic decay $\Sigma^0\rightarrow \Lambda+\gamma$ and the weak radiative decays of polarized hyperons, \textit{i.e.} decays of the type $Y\rightarrow B'+\gamma$ shown by the diagram in Fig. \ref{fig:decays EM WR}. We label the momenta of the particles as in Sec. \ref{sec:hyperon decays}.

\subsection{Electromagnetic decay}
The covariant matrix element describing the electromagnetic decay $Y\rightarrow B'+\gamma$ can be obtained by writing $\Mcov=\Mcov^{\mu}\phi^*_{\gamma,\mu}$, $\phi^{\mu}_{\gamma}$ being the polarization fourvector of the emitted photon, and expanding $\Mcov^{\mu}(p',q)$ in terms of Dirac matrices, $q=p-p'$ being the fourmomentum of the photon. Requiring parity invariance and electromagnetic current conservation leads to $\Mcov \propto -\i\,\sigma^{\mu\nu}\,q_{\nu}\,\phi^*_{\gamma,\mu}$, where $\sigma^{\mu\nu}=\i\,[\gamma^{\mu},\gamma^{\nu}]/2$. In the rest frame of $Y$, it gives the reduced matrix element
\begin{eqnarray}\label{eq: Mem}
    \Mhat_{\rm{EM}} \propto -\sigmabf \cdot [\Pphat\times \epsbf^*_{\gamma}],
\end{eqnarray}
where we have kept only the spin-dependent part of the matrix element. 
$\epsbf_{\gamma}$ is the vector amplitude of $\gamma$ in the Coulomb gauge.

The matrix element in Eq. (\ref{eq: Mem}) leads to the decay distribution
\begin{eqnarray}\label{eq:dN EM}
\nonumber    \frac{dN^{\EM}_{Y\rightarrow B'+\gamma}}{d\Omega_{\Pphat}} \propto\sum_{\epsbf_{\gamma}}\,\Tr_{B'}\left[\Mhat_{\rm{EM}}\,\rho(Y)\,\Mhat^{\dagger}_{\rm{EM}}\,\check{\rho}(B')\right],\\
\end{eqnarray}
where we have summed over the polarization states of the (not analysed) $\gamma$, and $\check{\rho}(B')$ is the acceptance matrix of $B'$.
Inserting Eq. (\ref{eq: Mem}) in Eq. (\ref{eq:dN EM}), the angular distribution of $B'$ is obtained and taking $\check{\rho}(B')=1_{2\times 2}$. We get $dN^{\EM}_{Y\rightarrow B'+\gamma}/d\Omega_{\Pphat}=(4\pi)^{-1}$, hence an isotropic angular distribution in the rest frame of $Y$.

If $B'$ is a hyperon to be also decayed, it's spin density matrix is needed and can be read off from Eq. (\ref{eq:dN EM}). We obtain
\begin{eqnarray}\label{eq:rho(B') EM}
\nonumber    \rho(B') &=& \frac{\sum_{\epsbf_{\gamma}}\, \Mhat_{\rm{EM}}\,\rho(Y)\,\Mhat^{\dagger}_{\rm{EM}}}{\Tr[\dots]} \\
&=& \frac{1}{2}\,\bigg[1 - \left(\SY\cdot\Pphat\right)\,\sigmabf\cdot\Pphat \bigg]
\end{eqnarray}
where the second line is obtained by using Eq. (\ref{eq: Mem}). The polarization vector of $B'$ is thus $\SBp=-(\SY\cdot\Pphat)\,\Pphat$, in agreement with helicity conservation in the decay process.

Once the decay of $Y$ has been simulated, according to the Collins-Knowles recipe \cite{Collins:1987cp,Knowles:1988vs}, it returns the acceptance matrix (also named decay matrix) $\check{\rho}(Y)=1+\sigmabf\cdot\SYcheck$, where $\SYcheck$ is the acceptance polarization vector of $Y$. This is necessary to propagate the information about the relative orientation of the decay products of $Y$ back to the emission vertex of $Y$, and to continue the propagation of the spin correlations along the fragmentation chain. The acceptance matrix can be deduced from Eq. (\ref{eq:dN EM}) interpreting the decay distribution as $\Tr[\rho(Y)\,\rhocheck(Y)]$. We obtain the not-normalized acceptance matrix
\begin{eqnarray}\label{eq:D_EM}
\nonumber    \check{\rho}(Y) &=& \sum_{\epsbf_{\gamma}}\Mhat_{\EM}^{\dagger}\,\check{\rho}(B')\,\Mhat_{\EM}
= 1 - \left(\SBpcheck\cdot\Pphat\right)\,\sigmabf\cdot\Pphat,\\
\end{eqnarray}
evaluated at the generated value of $\Pphat$. The acceptance polarization vector of $Y$ is therefore $\SYcheck=-\left(\SBpcheck\cdot\Pphat\right)\,\Pphat$, where $\SBpcheck$ is the acceptance polarization vector stemming from the decay of an unstable $B'$.
If the decay of $B'$ is not analysed, it is $\SYcheck=\textbf{0}$.

\subsection{Weak radiative decay}
To describe the weak radiative decay $Y\rightarrow B'+\gamma$, we take the covariant matrix element $\Mhat_{cov.}=F_1\,\Gamma^{\mu}\phi_{\gamma,\mu}^*+iF_2\sigma^{\mu\nu}q_{\mu}\phi_{\gamma,\nu}^*+F_3\gamma^{\mu}\gamma_5\phi_{\gamma,\mu}^*$, where $F_1$, $F_2$ and $F_3$ are form factors \cite{ParticleDataGroup:2022pth}. In the rest frame of $Y$, the covariant matrix element reduces to
\begin{eqnarray}\label{eq:M RD}
    \Mhat_{\RD} \propto -iF_{M}\,\sigmabf\cdot(\Pphat\times\epsbf_{\gamma}^*) + G\,\sigmabf\cdot\epsbf^*_{\gamma},
\end{eqnarray}
where the relevant form factors are defined to be $F_M=(M-M')\,[F_2-F_1/(M+M')]$ and $G=F_3$ \cite{ParticleDataGroup:2022pth}, $\Pphat$ is the direction of the momentum of the decay baryon $B'$, and $\epsbf_{\gamma}$ is the vector amplitude of the photon in the Coulomb gauge.

The angular distribution of $B'$ can be calculated by Eq. (\ref{eq:dN EM}) by the substitution $\Mhat_{\EM}\rightarrow \Mhat_{\RD}$, and by summing over the polarization states of $B'$. We obtain
\begin{eqnarray}\label{eq:dN RD}
    \frac{dN_{Y\rightarrow B'+\gamma}^{\RD}}{d\Omega_{\Pphat}} = \frac{1}{4\pi}\left(1+\alpha_{Y,\gamma}\,\SB\cdot\Pphat\right),
\end{eqnarray}
where the parameter $\alpha_{Y,\gamma}=2\Re(F_M^*\,G)/[|F_M|^2+|G|^2]$ depends on the decaying hyperon and can be found in the Ref. \cite{ParticleDataGroup:2022pth}.

Substituting  $\Mhat_{\EM}$ in Eq. (\ref{eq:rho(B') EM}) with $\Mhat_{\RD}$ in Eq. (\ref{eq:M RD}), one can obtain the spin density matrix of $B'$. We obtain
\begin{eqnarray}\label{eq:rho(B') RD}
    \rho(B') &=& \frac{\sum_{\epsbf_{\gamma}}\, \Mhat_{\RD}\,\rho(Y)\,\Mhat^{\dagger}_{\RD}}{\Tr[\dots]} \\
\nonumber &=& \frac{1}{2}\,\bigg[1-\sigmabf\cdot \Pphat\frac{\alpha_{Y,\gamma}+\SY\cdot\Pphat}{1+\alpha_{Y,\gamma}\SY\cdot\Pphat}\bigg].
\end{eqnarray}
The spin density matrix of $B'$ depends thus only on the combination of form factors given by the measured parameter $\alpha_{Y,\gamma}$. Moreover, $B'$ is polarized even if $Y$ is unpolarized, due to parity non-conservation. For $\SY=\textbf{0}$, it is $\SBp=-\alpha_{Y,\gamma}\Pphat$.

The acceptance matrix associated to the radiative decay of $Y$ can be obtained as in Eq. (\ref{eq:D_EM}) using $\Mhat_{\RD}$ in Eq. (\ref{eq:M RD}) instead of $\Mhat_{\EM}$. We obtain
\begin{eqnarray}\label{eq:D RD}
\nonumber    \check{\rho}(Y) &=& \frac{1}{2}\,\bigg[1+\sigmabf\cdot \Pphat\frac{\alpha_{Y,\gamma}-\SBpcheck\cdot\Pphat}{1-\alpha_{Y,\gamma}\SBpcheck\cdot\Pphat}\bigg].
\end{eqnarray}
If the decay of $B'$ is not analyzed, the acceptance matrix of $B$ is characterized by the polarization vector $\SYcheck=\alpha_{Y,\gamma}\,\Pphat$. 
\end{appendix}
\bibliography{bibliography}

\end{document}